\documentclass[3p]{elsarticle}
\pdfoutput=1
\usepackage[unicode=true, bookmarks=false,bookmarksnumbered=true, bookmarksopen=true, breaklinks=false, backref=false, pagebackref=false, colorlinks=true]{hyperref}
\hypersetup{pdftitle={}, pdfauthor={Nuno Cardoso and Pedro Bicudo}, linkcolor=black, citecolor=black, urlcolor=black, filecolor=black, pdfpagelayout=OneColumn, pdfnewwindow=true, pdfstartview=XYZ, plainpages=false}
\usepackage{color}
\usepackage{amsmath}
\usepackage{amssymb}
\usepackage{braket}
\usepackage[T1]{fontenc}
\usepackage{url}
\usepackage[english]{babel}
\usepackage{dcolumn}
\usepackage{bm}
\usepackage{subfig}
\usepackage{float}
\usepackage{url}
\usepackage{esint}
\usepackage{dsfont}
\usepackage{listings}
\newcommand{\Tr}{\mbox{\rm Tr}}
\newcommand{\ReC}{\mbox{\rm Re}}
\newcommand\T{\rule{0pt}{2.6ex}}
\newcommand\B{\rule[-1.2ex]{0pt}{0pt}}

\usepackage{verbatim}

\journal{Computer Physics Communications}

\begin{document}
\begin{frontmatter}

\title{Generating SU($N_c$) pure gauge lattice QCD configurations \\
on GPUs with CUDA}
\author{Nuno Cardoso\corref{cor1}}
\ead{nuno.cardoso@ist.utl.pt}
\author{Pedro Bicudo}
\ead{bicudo@ist.utl.pt}
\address{CFTP, Departamento de F\'{i}sica, Instituto Superior T\'{e}cnico, Universidade T\'{e}cnica de Lisboa, Av. Rovisco Pais, 1049-001 Lisbon, Portugal}

\cortext[cor1]{Corresponding author}

\begin{abstract}
The starting point of any lattice QCD computation is the generation of a Markov chain of gauge field configurations. Due to the large number of lattice links and due to the matrix multiplications, generating SU($N_c$) lattice QCD configurations is a highly demanding computational task,  requiring advanced computer parallel architectures such as clusters of several Central Processing Units (CPUs) or Graphics Processing Units (GPUs). 
In this paper we present and explore the performance of CUDA codes for NVIDIA GPUs to generate SU($N_c$) lattice QCD pure gauge configurations. Our implementation in one GPU uses CUDA and in multiple GPUs uses OpenMP and CUDA.
We present optimized CUDA codes for SU(2), SU(3) and SU(4). We also show a generic SU($N_c$) code for $N_c\,\geq 4$ and compare it with the optimized version of SU(4). Our codes are publicly available for free use by the lattice QCD community.
\end{abstract}


\end{frontmatter}

\section{Introduction}

Generating SU($N_c$) lattice configurations is a highly demanding computational task and requires advanced computer architectures such as clusters of several Central Processing Units (CPUs) or Graphics Processing Units (GPUs). 
Compared with CPU clusters, GPUs are easier to access and maintain, as they can run on a local desktop computer. 

CUDA (Compute Unified Device Architecture) is NVIDIA's parallel computing architecture which enables dramatic increases in performance computing using GPUs.
Since 2007, the year when NVIDIA released CUDA for GPU computing as a language extension to C, CUDA has become a standard tool in the scientific community. The CUDA architecture also supports standard languages, such as C and Fortran, and APIs for GPU computing, such as OpenCL and DirectCompute.
With GPUs, many scientific problems can now be addressed without the need to use a cluster of CPUs or by using clusters of GPUs in which the computation time can be reduced significantly, for example \cite{Gross20111638, Zonoozi:2010bi, Hassan:2011nb, Kanzaki:2011, JCC:JCC20829, Tomov2010232, Cardoso:2010di,Cardoso:2012pv}.

The most successful theories of elementary particle physics are the gauge theories. They are renormalizable 
\cite{'tHooft:1971rn}
and have Lie groups as internal gauge symmetries. 
The  special unitary groups SU($N_c$) are cornerstones of the the Standard Model of particle physics, especially SU(2) together with U(1)  in the electroweak interaction and SU(3) in quantum chromodynamics (QCD), the theory of the strong interaction.  Moreover, since QCD is not yet fully understood, it is relevant to study the effect of changing the number of colors, $N_c$, studying other SU($N_c$) groups.   
In particular, since the seminal works of 't Hooft 
\cite{'tHooft:1973jz},
Witten
\cite{Witten:1980sp}
and Creutz
\cite{Creutz:1980hb,Creutz:1980zy,Creutz:1981gza}, 
the large $N_c$ limit of QCD has been explored in great detail \cite{Teper:2009uf,Panero:2009tv,Lohmayer:2011ns,Lohmayer:2011nq,Bursa:2012ab,Lohmayer:2012ue,Mykkanen:2012ri}.  


However, the only existing approach to study SU($N_c$) gauge theories namely in strong interactions and beyond in a non-perturbative way is the lattice field theory.  
Based on the path integral formalism and in statistical mechanics methods, the observables are evaluated numerically using Monte Carlo simulations. The starting point of any lattice QCD computation is the generation of a Markov chain of gauge field configurations. The configuration generation, due to the large number of lattice links and to the matrix multiplications, is computationally expensive. 
Thus,  in the lattice community many groups
\cite{Egri:2006zm, Clark:2009wm, Hayakawa:2010gm, Kim:2010br, Bonati:2010qu, Walk:2010ut,Babich:2010mu, Chiu:2011rc,Alexandru:2011ee, Winter:2011an, Alexandru:2011sc, Bonati:2011dv, Babich:2011np} 
are already using GPUs to generate lattice QCD configurations. They are specialized in SU(3) and in using the GPUs for the full Lagrangian description, i.e., gluons together with dynamical quarks.   

Here we describe and study the performance of our configuration generation codes in pure gauge lattice QCD.
Pure lattice gauge theory does not include the full Lagrangian description, i.e., the quarks are fixed and therefore this approach is also denominated as quenched approximation. In the quenched lattice approach, the required computational power to generate pure gauge configurations, although quite intensive, is one or two orders of magnitude less demanding than the full lattice approach. Although quenched QCD is a simplification of QCD, there are still many problems that remain to be solved in that approach. In particular it is computationally feasible to study quenched QCD with a larger $N_c$.

Recently 
\cite{Cardoso:2010di},
we addressed the GPU computational power necessary to generate pure gauge SU(2) configurations. We showed that a server with a single CPU commanding  a few GPUs is quite efficient to generate gauge SU(2) configurations with codes including the Open Multi-Processing (OpenMP) and CUDA libraries. Here we extend our previous code for SU(2) to SU(3), SU(4) and to a generic SU($N_c$) code.

This paper is divided into 5 sections. In section 2, we present a brief description on
how to generate lattice SU($N_c$) configurations with $N_c\geq2$ and in section 3 we show the implementation in one GPU using CUDA or in multiple GPUs using OpenMP and CUDA. In section 4, we present the GPU performance in single and double precision using one, two and three GPUs for different lattice volumes. Finally, in section 5, we conclude.

\section{Lattice Gauge Theory}

Gauge theories can be addressed by lattice field theory in a non-perturbative approximation scheme, based on the path integral formalism in which space-time is discretized on a 4-dimensional hypercubic lattice. In a lattice, the fields representing the fermions are defined in lattice sites, while the gauge fields are represented by link variables $U_\mu(x)$ connecting neighboring sites. In this work, we employ the pure gauge quenched approximation.

\begin{figure}
\begin{centering}
    \includegraphics[width=10cm]{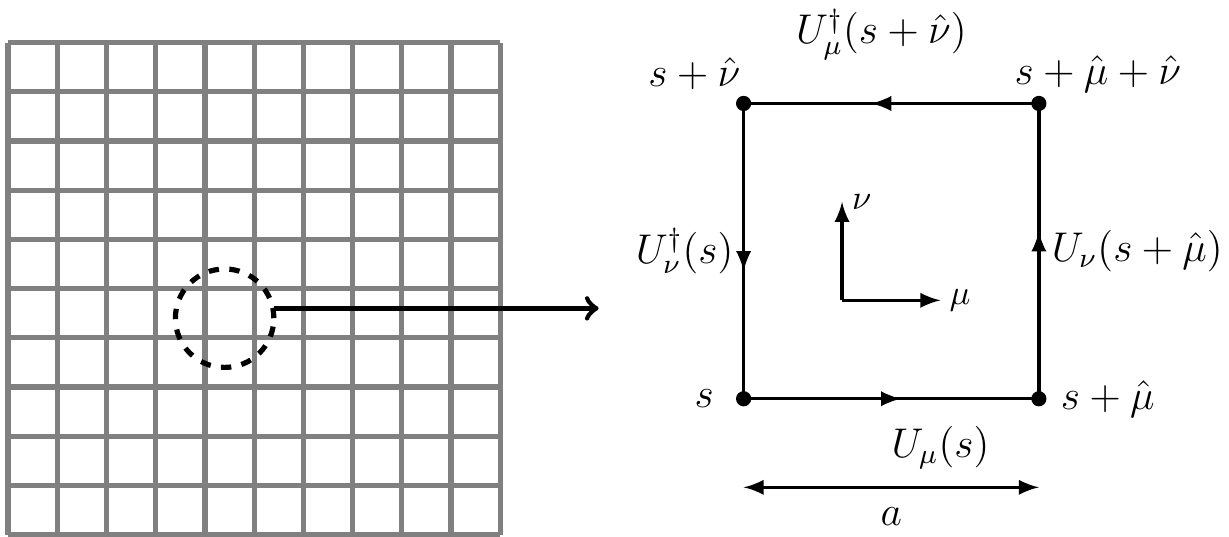}
\par\end{centering}
\caption{Plaquette $P_{\mu\nu}(s)$.}
\label{fig:plaq}
\end{figure}

In quenched lattice QCD, quantities in the form of a path integral are transformed to Euclidean space-time, and are evaluated numerically using Monte Carlo simulations allowing us to use statistical mechanics methods.
The partition function in Euclidean space-time, in the quenched approximation, is given by
\begin{equation}
Z=\int \mathcal{D}[U]\exp\left(- S_G[U]   \right)\, ,
\end{equation}
where the integration measure for the link variables is the product measure
\begin{equation}
\int \mathcal{D}[U]=\prod_s\prod_{\mu=0}^3\int dU_\mu(s)\, ,
\end{equation}
and $S_G[U]$ is the gauge field action. Here we use the Wilson gauge action, defined by
\begin{equation}
S_G[U] = \frac{\beta}{N_c} \sum_s\sum_{\mu<\nu} \ReC\Tr\left[ 1 - P_{\mu\nu}(s)  \right]
\end{equation}
where $\beta=2 \, N_c/g^2$ is the inverse of the gauge coupling and $ P_{\mu\nu}(s)$ is the plaquette, Fig. \ref{fig:plaq}, defined as
\begin{equation}
P_{\mu\nu}(s) = U_\mu(s)U_\nu(s+\hat{\mu})U^\dagger_\mu(s+\hat{\nu})U^\dagger_\nu(s)\, .
\end{equation}
Physical observables are obtained calculating the expectation
value,
\begin{equation}
\Braket{\mathcal{O}}=\frac{1}{Z}\int \mathcal{D}[U]\exp\left(- S_G[U]   \right) \mathcal{O}[U]\, ,
\end{equation}
where $\mathcal{O}$ is given as a combination of operators expressed in terms of time-ordered products of fields.

We set up our pure gauge lattice on a 4-dimensional hypercubic lattice with spacing
$a$, time-like extent $T=N_t$  and spatial size $L=N_x\times N_y\times N_z=N_\sigma$, with periodic boundary conditions. 
The lattice QCD Monte Carlo simulations approximate the integral by an average of the observable evaluated on N sample gauge field configurations with distribution probability $\propto \exp (-S_G[U])$. The sequence of configurations generated by Monte Carlo algorithms produces a Markov chain.
Each Monte Carlo step consists in visiting and updating all gauge links in the lattice. There are different algorithms to update a gauge link, such as Metropolis and heatbath. We will use the heatbath method, since it is more efficient than the Metropolis algorithm.

The gauge field variables of an SU($N_c$) gauge group are represented on the lattice links by complex $N_c\times N_c$ matrices. Using the unitarity of the group elements, we may use a minimal set of parameters equal to the number of generators of the group. However, in practical calculations, it is more convenient to use a redundant parameterization of the gauge group. For example, the SU(2) group can be represented by four real numbers instead of using $2\times2$ complex matrices, since a $ 2 \times 2$ matrix parameterized  with,
\begin{equation}
U=a_0 \mathds{1} + i\,\mathbf{a}\cdot \sigma 
\end{equation}
with
\begin{equation}
a^2=a^2_0 + \mathbf{a}^2 = 1 
\end{equation}
is an SU(2) matrix, and vice versa
\begin{equation}
\Tr\, U=2\,a_0,\quad \quad U U^{\dagger}=U^{\dagger}U=\mathds{1},\quad \quad \det U = 1 \ .
\end{equation}

Generating SU($N_c$) lattice configurations is a highly demanding computational task. Most of the computational time is spent in updating the gauge links. We start by describing the update link method for the SU(2) group, since this is the basis for the groups with $N_c\geq3$.
In order to update a particular link in SU(2), \cite{Creutz:1980zw,Creutz:1980zj}, we need only to consider the contribution to the action from the six plaquettes containing that link, the staple $\Sigma$,
\begin{equation}
\Sigma = \sum_{\mu\neq\nu}(U_{x,\nu}U_{x+\hat{\nu},\mu}U^\dagger_{x+\hat{\mu},\nu} + U^\dagger_{x-\hat{\nu},\nu}U_{x-\hat{\nu},\mu}U_{x-\hat{\nu}+\hat{\mu},\nu})\ .	
\end{equation}
The distribution to be generated for every single link is given by
\begin{equation}
dP(U)\propto \exp \left[\frac{1}{2} \beta \Tr(U\Sigma)\right]\ .
\end{equation}
Applying a useful property of SU(2) elements, that any sum of them is proportional to another SU(2) element $\tilde{U}$,
\begin{equation}
\tilde{U}=\frac{\Sigma}{\sqrt{\det \Sigma}}=\frac{\Sigma}{k}\ .
\end{equation}
and using the invariance of the group measure, we obtain
\begin{equation}
dP\left(U\tilde{U}^{-1}\right) \propto \exp\left[\frac{1}{2}\beta k \Tr U\right]dU = \exp\left[\beta k a_0\right] \frac{1}{2\pi^2}\delta\left(a^2-1\right)d^4 a\ .
\end{equation}
Thus, we need to generate $a_0\in [-1,1]$ with distribution,
\begin{equation}
P\left(a_0\right) \propto \sqrt{1-a_0^2} \exp\left(\beta k a_0\right)\ .
\end{equation}
and the components of $\mathbf{a}$ are generated randomly on the 3D unit sphere in a 4-dimensional space with exponential weighting along the $a_0$ direction.
Once the $a_0$ and $\mathbf{a}$ are obtained in this way, the new link is updated,
\begin{equation}
U'=U\tilde{U}^{-1}\ .
\end{equation}

Although for SU($N_c$) with $N_c\geq3$ there is no heatbath algorithm which directly produces SU($N_c$) link variables, we can apply a pseudo-heatbath method, also known as the Cabibbo-Marinari algorithm \cite{Cabibbo:1982zn}, for the SU(2) subgroups of SU($N_c$).
The procedure to update a link for $N_c\geq3$ is
\begin{enumerate}
\item calculate the staple, $\Sigma$;
\item calculate the $U\Sigma^\dagger$;
\item select a set of SU(2) subgroups of SU($N_c$) from the previous result, such that there is no subset of SU($N_c$) left invariant under left multiplication, except the whole group;
\item although the minimal set may involve only $N_c-1$ subgroups, here we decide to use the complete set, i.e., $N_c(N_c-1)/2$ subgroups;
\item the update of a given link is done in $k$ steps, $k=1,...,N_c(N_c-1)/2$. In each step is generated a member of $A_k\in$ SU(2)$_k$. Then the current link at that step is obtained by multiplying the link obtained in the last step by $A_k$,
	\begin{equation}
	U^k=A_k U^{k-1}\ .
	\end{equation}
\end{enumerate}

For example, although in $N_c=3$ two subgroups would be sufficient to cover the whole group space, for symmetry reasons we will use all subgroups.
In $N_c=3$, the three subgroups of SU(3), $S_{ij}$ with $1\leq i<j\leq 3$ matrices are constructed as
\begin{equation}
\begin{array}{ccccc}
\begin{pmatrix}a_{11} & a_{12} & 0\\
a_{21} & a_{22} & 0\\
0 & 0 & 1
\end{pmatrix} &  & \begin{pmatrix}1 & 0 & 0\\
0 & a_{11} & a_{12}\\
0 & a_{21} & a_{22}
\end{pmatrix} &  & \begin{pmatrix}a_{11} & 0 & a_{12}\\
0 & 1 & 0\\
a_{21} & 0 & a_{22}
\end{pmatrix}\end{array}
\end{equation}
with $a_{ij}\in$ SU(2).
Each of the $N_c(N_c-1)/2$ subgroups is determined using the heatbath for SU(2) as already discussed.

In order to accelerate the decorrelation of subsequent lattice configurations, we can employ the over-relaxation algorithm, which in SU(2) is defined by
\begin{equation}
U_\text{new} = \frac{\Sigma^\dagger}{\left|\Sigma\right|} U^\dagger_\text{old}\frac{\Sigma^\dagger}{\left|\Sigma\right|}\ ,
\end{equation}
with $|\Sigma| = \sqrt{\det\Sigma}$.
Although for SU(2) group this is straightforward, for the SU($N_c$) with $N_c\geq3$ this is not the case. However, the method is similar to the pseudo-heatbath method and we can use the above equation for each of the SU(2) subgroups of SU($N_c$).

Because of the accumulation of rounding errors in the multiplications of the group elements, the matrices have to be regularly projected to unitarity. This step in the algorithm is called re-unitarization. Re-unitarization for $N_c=2$ is done by normalizing the first row and then reconstructing the second row from the first. For $N_c\geq3$, this is done using the well-known Gram-Schmidt method for building an orthonormal basis element in vector spaces. For $N_c>3$, after the Gram-Schmidt method, we need to multiply the last row with a phase to make the determinant equal to one.

\section{Implementation}

In this section, we discuss the parallelization scheme for generating pure gauge SU($N_c$) lattice configurations on GPUs using CUDA, with optimized codes for $N_c=2,3,4$ and a generic code for $N_c \ge 4$.

We implement and test our codes for generating pure gauge lattice configurations in CUDA version 3.2. For our 4-dimensional lattice, we address one thread per lattice site.  Version 3.2 only supports thread blocks up to 3D and grids up to 2D, and the lattice needs four indexes.
Therefore we compare 1D thread blocks where we reconstruct all the indexes on the fly with 3D thread blocks, one for t, one for z and one for both x and y and then reconstruct the other index inside the kernel.

Since the grid can have only 65535 thread blocks per dimension, for a large lattice volume this number is insufficient, i.e., using one thread per site, we can only have  $(\text{lattice volume}) / (\text{number of threads}) \leq 65535$. 
We put most of the constants needed by the GPU, like the number of points in the lattice, in the GPU constant memory using cudaMemcpyToSymbol.  
For the lattice array we cannot use a 4-dimensional array to store the lattice in CUDA. Therefore, we use a 1D array with size $N_x\times N_y\times N_z\times N_t\times Dim$, with $Dim=4$.

For $N_c=2$, we only need four floating point numbers per link instead of having a $2\times2$ complex matrix, therefore we use an array of structures. Each array position is a float4/double4 structure to store the generators of SU(2) (a0, a1, a2 and a3). 
When accessing the global memory, 128-bit words give fully coalesced memory transactions. Although in single precision we can use a float4 (128-bit word) array to store all the generators of SU(2), this is not the case in double precision. Using a double4 format does not give fully coalesced memory transactions since it is a 256-bit word, whereas the double2 format is a 128-bit word and gives fully coalesced memory transactions.
Therefore, we also implemented an array of double2. First we store the first two generators, a0 and a1, for all the lattice size and then the last two generators, a2 and a3.

For $N_c=3 \text{ and }4$, we tested an array of structures (AOS) and a structure of arrays (SOA). Since each element of SU(3) and SU(4) is a complex number, we use the float2/double2 CUDA vector types. Therefore, in a AOS each array index is a structure with $N_c\times N_c$ float2/double2 elements. The SOA structure is composed by $N_c\times N_c$
arrays of type float2/double2. To select single or double precision, we have implemented templates in the code.

In the SU(3) case, we have also implemented an SOA with 12 arrays of type float2/double2. As discussed in section 2, we can use a minimal set of parameters equal to the number of generators of the group. In SU(3), the minimum is 8 parameters, however this is not numerically stable, \cite{Clark:2009wm,Babich:2010mu}. But if we use 12 parameters instead of 8 parameters, storing only the first two rows and reconstructing the third row on the fly, the truncation errors are smaller and we can reduce memory traffic.

\begin{table}[!htb]
\begin{center}
\begin{tabular}{|c|c|c|c|c|}
\hline
\T\B \textbf{Kernel} & PHB and OVR & REU & PLAQ\\
\T\B per thread & per link & per site & per site\\ \hline
\hline
\T\B \textbf{SU(2) 4 reals} & 76 &  none & 96\\ \hline
\T\B Single/Double (bytes) & 304/608 & none & 384/768 \\
\hline
\hline
\T\B \textbf{SU(3) 18 reals} & 342 &  72 & 432 \\ \hline
\T\B Single/Double (bytes) & 1368/2736 & 288/576 & 1728/3456  \\ \hline
\hline
\T\B \textbf{SU(3) 12 reals} & 228 &  48 & 288 \\ \hline
\T\B Single/Double (bytes) & 912/1824 & 192/384 & 1152/2304 \\ \hline
\hline
\T\B \textbf{SU(4) 32 reals} & 608 & 128 & 768  \\ \hline
\T\B Single/Double (bytes) & 2431/4864 &  512/1024 & 3072/6144 \\ \hline
\hline
\end{tabular}
\end{center}
\caption{Kernel memory loads per thread for pseudo-heatbath kernel (PHB), over-relaxation kernel (OVR), re-unitarization kernel (REU) and plaquette kernel (PLAQ).}
\label{tab:memload}
\end{table}

\lstset{emph={double2, float2,double4, float4,__host__, __device__,cudaMalloc,CUDA_SAFE_CALL}, emphstyle={\color{black}\textbf},emph={[2]}, emphstyle={[2]\color{black}}}
We now detail the structures used for the SU($N_c$) configuration codes with different $N_c$. Since the $N_c \ge 4$  implementation is similar to the one of the SU(3) code, we show in more detail the SU(3) implementation. The difference between the structures of these codes is the number of elements, say, 18/12 real numbers for the SU(3) implementation with full and 12 real numbers matrix parameterization, and 32 real numbers for the SU(4) implementation.

\begin{figure}[!htb]
\begin{centering}
    \subfloat[\label{fig:Grid}]{
\begin{centering}
    \includegraphics[width=4.0cm]{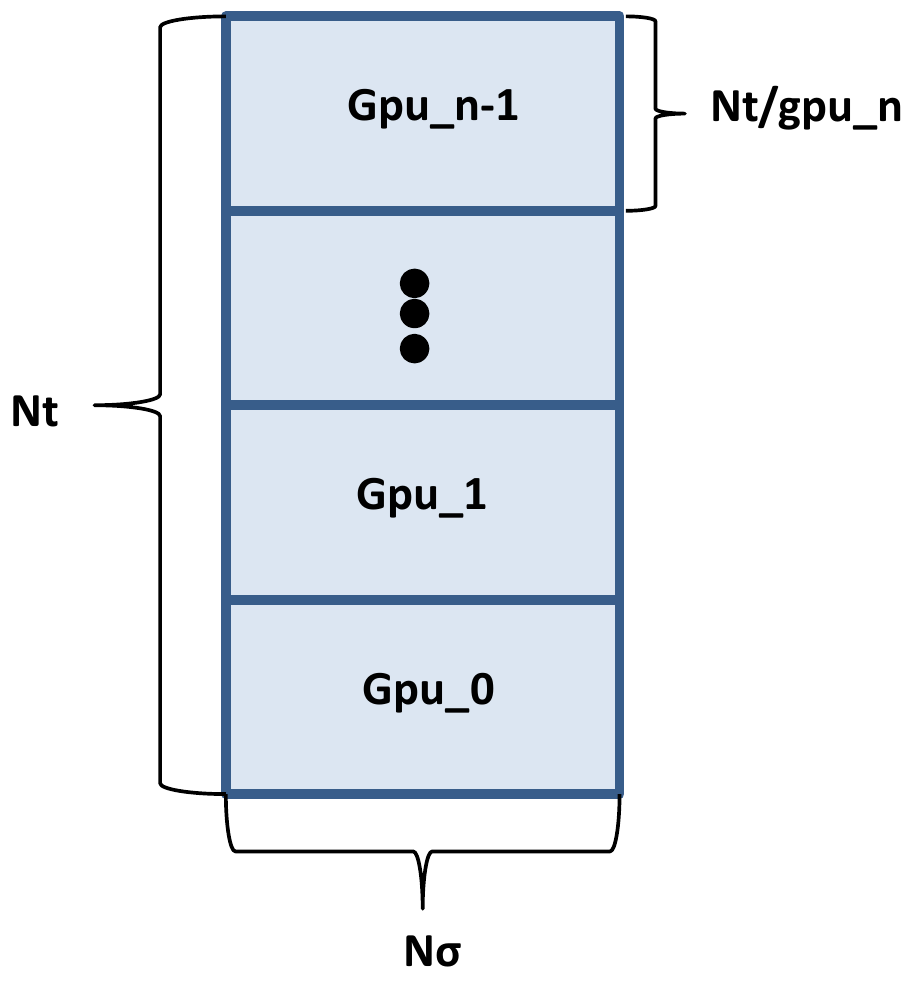}
\par\end{centering}}
    \subfloat[\label{fig:Grid2}]{
\begin{centering}
    \includegraphics[width=4cm]{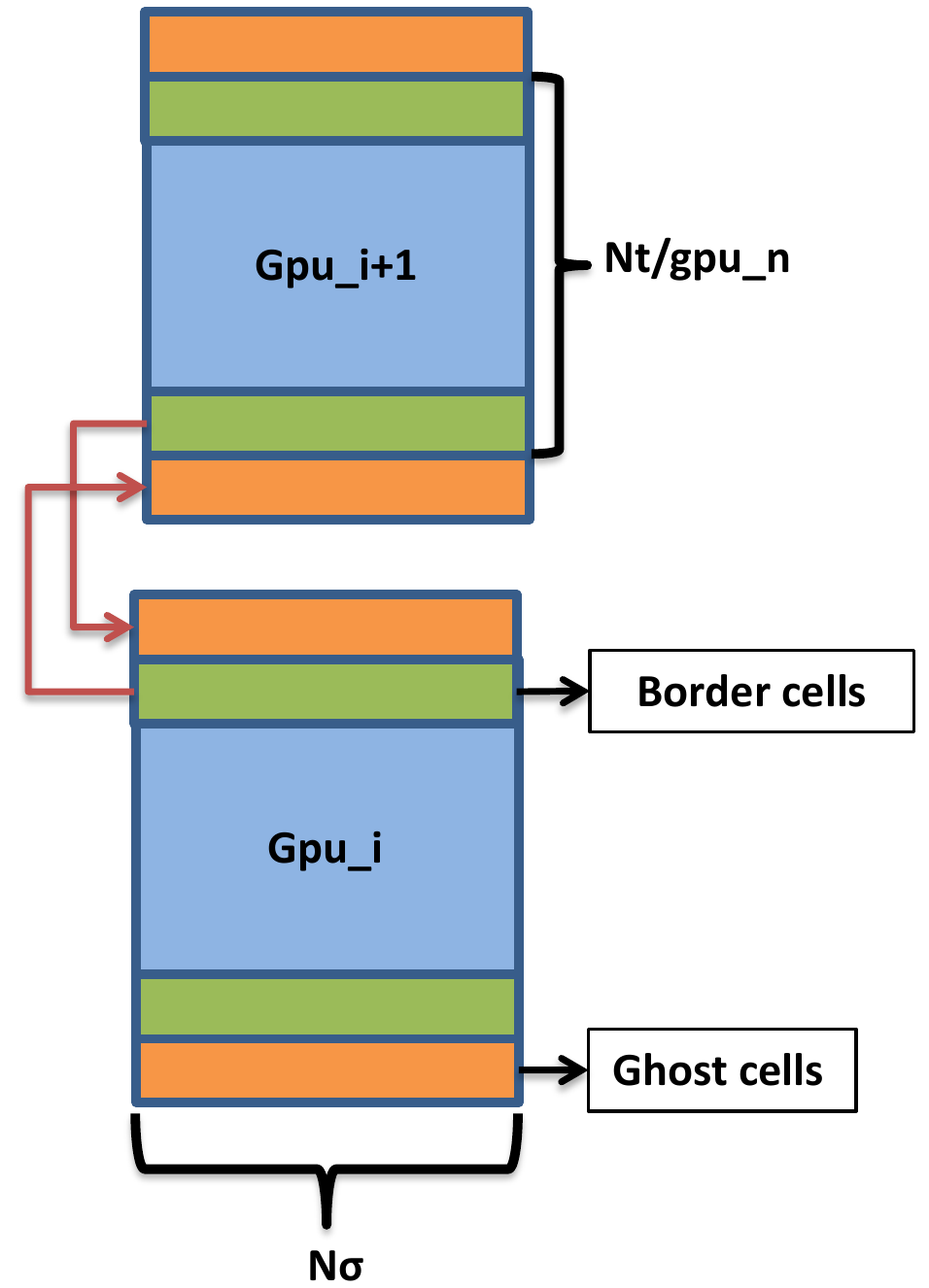}
\par\end{centering}}
\par\end{centering}
    \caption{Schematic view of the lattice array handled by each GPU.}
    \label{OpenMP_grid}
\end{figure}

For generating pure gauge lattice configurations, we implemented six kernels. 
Notice that in the heatbath and over-relaxation methods, since we need to calculate the change of the action for a Monte Carlo step by addressing the nearest neighbors links in the four space-time directions, we employ the chessboard method, calculating the new links separately by direction and by even/odd sites.
\begin{itemize}
\item Initialization of the array random number generator. We use the CURAND library of NVIDIA \cite{curandlib}.

\item Initialization of the lattice array. The initialization can be done with a random SU($N_c$) matrix (hot start) or with the identity matrix (cold start).

\item Link update by heatbath algorithm. Note that for $N_c\geq3$ this method is called the pseudo-heatbath algorithm, as discussed in section 2. This kernel must be called eight times, since this must be done by link direction and even and odd sites separately, because we need to calculate the staple at each link direction. 

\item Lattice over-relaxation. This kernel has to be implemented in the same way as the kernel to update each link by the heatbath method kernel, by link direction and even/odd sites separately.

\item Lattice re-unitarization, with the standard Gram-Schmidt technique, implemented only for SU($N_c$) with $N_c\ge$ 3 CUDA codes.

\item Plaquette at each lattice site. The sum over all the lattice sites is done with the parallel reduction code in the NVIDIA GPU Computing SDK package \cite{parallelreduction}, which is already a fully optimized code.
\end{itemize}
In Table \ref{tab:memload}, we summarize the memory loads per thread by kernel. The lattice SU($N_c$) is very memory traffic consuming. In the pseudo-heatbath and over-relaxation methods, to update a single link, it is necessary to copy from the lattice array memory 18 links, which make the staple, plus the link to be updated.

We now address the multi-GPU approach using CUDA and OpenMP. In order to use and control several GPUs on the same system, we need to have one CPU thread per GPU. The OpenMP allows us to do this when we have several GPUs on the same system. 
Therefore, we split the lattice along the temporal part among several GPUs, see Fig. \ref{fig:Grid}. The total lattice size in each GPU is now $N_\sigma\times4\times (N_t/(num.\, gpus) + 2)$, with four for the link direction and two for the neighboring sites (ghost cells), see Fig. \ref{fig:Grid2}. After each kernel, the border cells in each GPU need to be exchanged between each GPU. 
For this reason, the links at the borders of each sublattice have to be transferred from one GPU to the GPU handling the adjacent sublattice.
In order to exchange the border cells between GPUs it is necessary to copy these cells to CPU memory and then synchronize each CPU thread with the command \lstinline!#pragma omp barrier! before updating the GPU memory (ghost cells).

Since memory transfers between CPU and GPU are very slow compared with other
GPU memory and in order to maximize the GPU performance, we should only use this
feature when it is extremely necessary. Hence, we only use CPU/GPU memory transfers
in three cases: at the end of the kernel to perform the sum over all lattice sites (copy the final result to CPU memory, plaquette), when using multi-GPUs (exchange the border cells between GPUs) and file storing independently pure gauge lattice configurations.

\section{Results}

Here we present the benchmark results using two different GPU architectures (GT200 and Fermi), Table \ref{tab:nvidia_gpu_specs}, in generating pure gauge lattice configurations. We also compare the performance with two and three Fermi GPUs working in parallel in the same motherboard, using CUDA and OpenMP.

We compare the performance using the texture memory (tex) and using the L1 and L2 cache memories (cache). In the GT200 architecture, since it does not have L1 and L2 caches, 
the cache label in the figures corresponds to accessing directly the global memory. In the figures we use the notation "tex" when using
the texture memory and "cache" when not using the texture memory.

To test the performance of each kernel implemented for the SU(2), SU(3), SU(4) and generic SU($N_c$) codes, we use CUDA Profiler to measure the time spent for each kernel. For the SU(2) code, we perform 300 iterations, where each iteration consists of one heatbath step and one over-relaxation step. At the end of each step the plaquette is calculated. For the SU(3), SU(4) and generic SU($N_c$) codes, the procedure is the same. However, after the pseudo-heatbath and over-relaxation steps, a matrix re-unitarization is done.

For the multi-GPU part, we measure the total time to make a cold start to the system (all the links are initialized with the identity matrix) and perform 300/100/100 iterations of the SU(2)/SU(3)/SU(4) codes with one (pseudo-)heatbath and over-relaxation step, followed by link re-unitarization and at the end of each iteration the plaquette is calculated.
Note that we do not take into account the time for the initialization of the CURAND random number generator.

In Table \ref{tab:memload}, we show the memory loads for each kernel (heatbath/pseudo-heatbath, over-relaxation, plaquette and re-unitarization) in the SU(2), SU(3) and SU(4) codes.

Our code can be downloaded from the Portuguese Lattice QCD collaboration homepage \cite{ptqcd}.

\begin{table}[!htb]
\begin{centering}
\begin{tabular}{|c|c|c|}
\hline
\T\B NVIDIA Geforce GTX & 295 & 580\tabularnewline
\hline
\hline
\T\B Number of GPUs & 2 & 1\tabularnewline
\hline
\T\B CUDA Capability & 1.3 & 2.0\tabularnewline
\hline
\T\B Multiprocessors (MP) & 30 & 16\tabularnewline
\hline
\T\B Cores per MP & 8 & 32\tabularnewline
\hline
\T\B Number of cores & 2$\times$240 & 512\tabularnewline
\hline
\T Global memory & 1792 MB GDDR3 & 3072 MB\tabularnewline
\B & (896MB per GPU) & GDDR5\tabularnewline
\hline
\T\B Number of threads per block & 512 & 1024\tabularnewline
\hline
\T\B Registers per block & 16384 & 32768\tabularnewline
\hline
\T\B Shared memory (per SM) & 16KB & 48KB or 16KB\tabularnewline
\hline
\T\B L1 cache (per SM) & None & 16KB or 48KB \tabularnewline
\hline
\T\B L2 cache (chip wide) & None & 768KB \tabularnewline
\hline
\T\B Clock rate (GHz) & 1.37 & 1.57 \tabularnewline
\hline
\T\B Memory Bandwidth (GB/s) & 223.8 & 192.4\tabularnewline
\hline
\end{tabular}
\par\end{centering}
\caption{NVIDIA's graphics card specifications used in this work. Using OpenMP we also work with two 295 GTX  boards (4 GPUs in total) and three 580 GTX boards (3 GPUs in total).}
\label{tab:nvidia_gpu_specs}
\end{table}

\subsection{SU(2) CUDA performance}

When accessing the global memory, copying 128-bit words gives fully coalesced memory transactions. Although in single precision we can use a float4 (128-bit word) array to store all the SU(2) elements, this is not the case in double precision. Using a double4 format does not give fully coalesced memory transactions since it is a 256-bit word, whereas the double2 format is a 128-bit word and gives fully coalesced memory transactions.

\begin{figure}[!htb]
\begin{centering}
    \includegraphics[width=11cm]{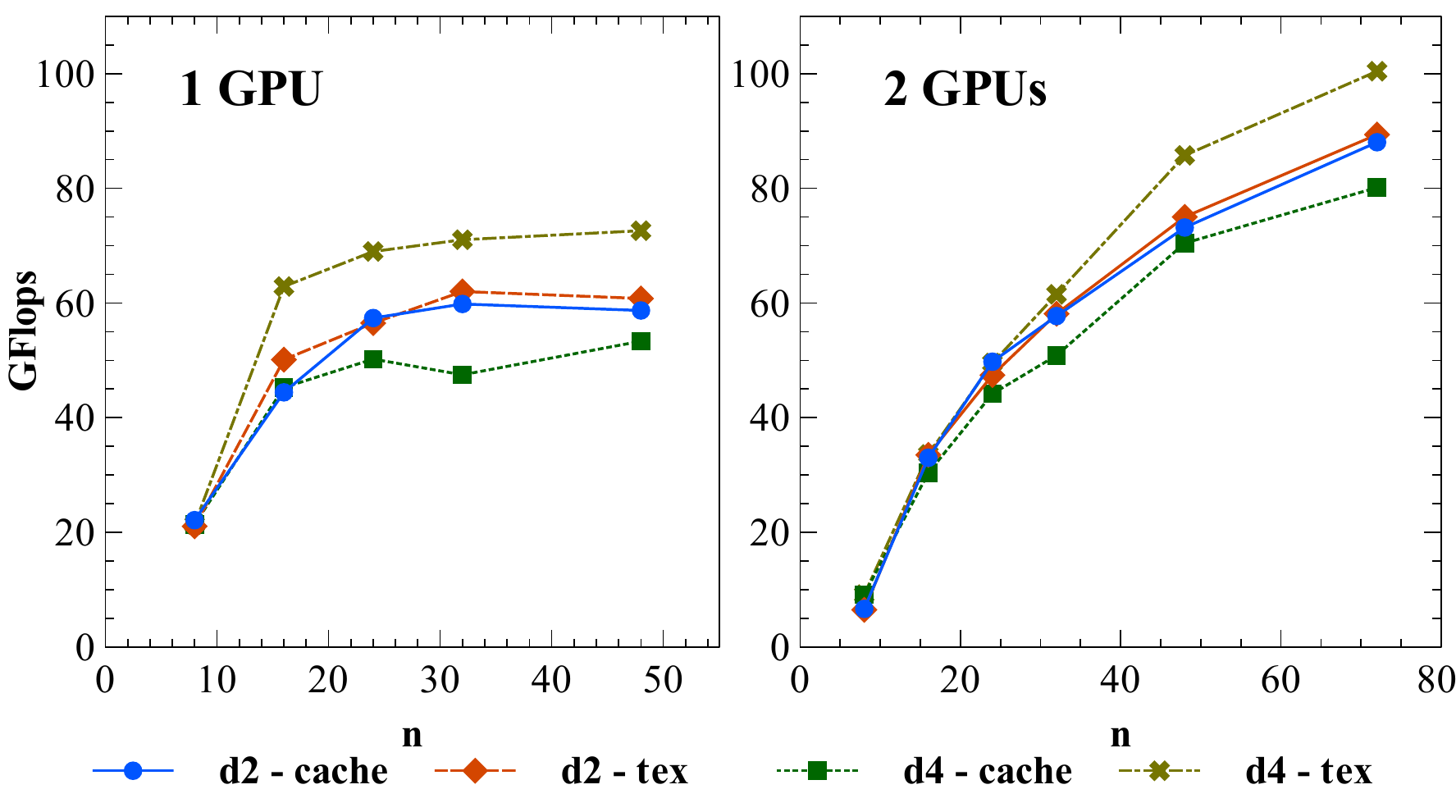}
\par\end{centering}
\caption{SU(2) CUDA performance in GFlops using one and two GTX 580 GPUs with CUDA and OpenMP in double precision using 3D thread blocks. d2 corresponds to a double2 array and d4 to a double4 array. "n" corresponds to the number of points in each lattice dimension, i.e., $n=N_x=N_y=N_z=N_t$. "tex" means using Texture memory and "cache" using cache memory.}
\label{fig:su2_perf_d2_d4}
\end{figure}

In Fig. \ref{fig:su2_perf_d2_d4}, we show the performance in double precision using a double4 array and a double2 array with one and two GPUs and 3D thread blocks. The best performance is obtained when using a double4 array and Texture memory. 
Nevertheless, using a double4 array and Texture memory we have achieved the highest performance using one or two GPUs.
However, if using the L1 and L2 caches, the maximum gain in performance of using a double2 array is 25\%/10\% using one/two GPUs compared with a double4 array. 
In the following performance results, we will use the double4 array.

\begin{figure}[!htb]
\begin{centering}
    \includegraphics[width=14cm]{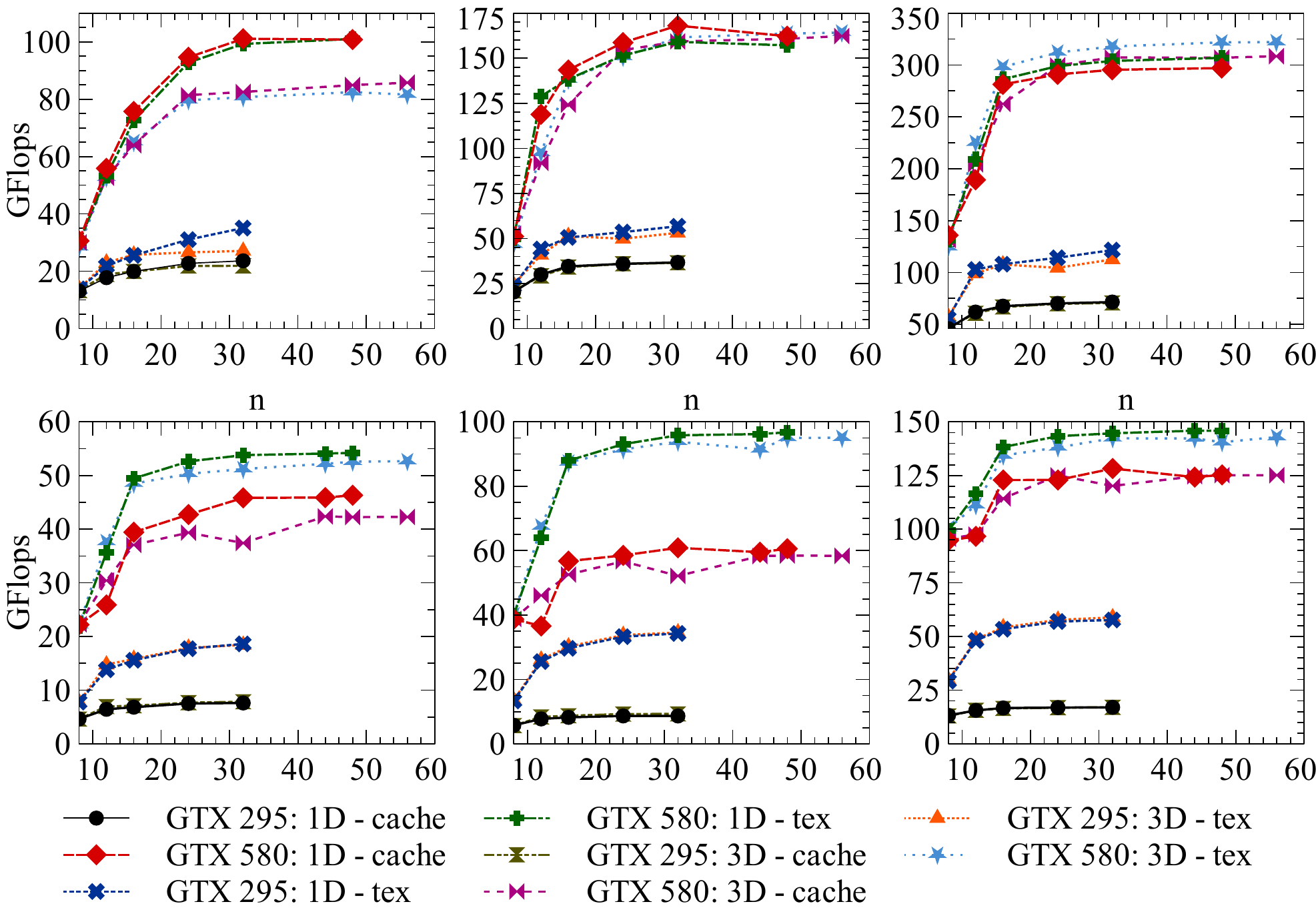}
\par\end{centering}
\caption{SU(2) CUDA performance. Left to right: heatbath, over-relaxation and plaquette kernels. The top line of graphics corresponds to performance in single precision and the bottom to double precision. "n" corresponds to the number of points in each lattice dimension, i.e., $n=N_x=N_y=N_z=N_t$.}
\label{fig:su2_perf}
\end{figure}

In Fig. \ref{fig:su2_perf}, we show the performance in GFlops as a function of the lattice volume, for each kernel in single GPU mode.
The heatbath and over-relaxation kernels are much slower than the plaquette kernel, updating a new link is the heavy part of the code. The performance of the over-relaxation kernel is higher than the heatbath kernel. Therefore, the link generation with some steps of over-relaxation increases the overall performance of the code, as it decreases the number of steps between configurations by decreasing the correlation time. In SU(2), it is common to use one step of heatbath followed by four steps of over-relaxation.

Since the heatbath and over-relaxation kernels have to update a new link using the staple, these kernels have to be called in eight steps, i.e., to perform a full lattice update and avoid a possible new neighboring link update, these kernels have to perform an independent update by link direction and by even/odd sites.

Note the performance in double precision is almost one half of the performance in single precision.

The best performance is obtained for 1D thread blocks as expected. When the lattice volume is not divisible by the number of threads, we have to add one more thread block which is not fully occupied. However when using 3D thread blocks this number can be higher. On the other hand, using 1D thread blocks, the total number of threads allowed is less than using 3D thread blocks, which limits the lattice size. 
To overcome this limitation, we may have to call the kernel several times in order to visit all the remaining sites, but this was not implemented in the code. As we mentioned in the previous section, the use of 1D thread blocks can only accommodate 65535 thread blocks, which can be insufficient when using large lattice volumes. For example, using 256 threads per block, we can have up to 256 thread blocks in a 1D grid and therefore the lattice volume must be less than $64^4$.
Moreover, the performance with 1D thread blocks is practically the same in double precision and higher than 3D thread blocks only in single precision. 

\begin{figure}[!htb]
\begin{centering}
    \subfloat[\label{fig:mgpu_su2_580_sp}]{
\begin{centering}
    \includegraphics[height=5cm]{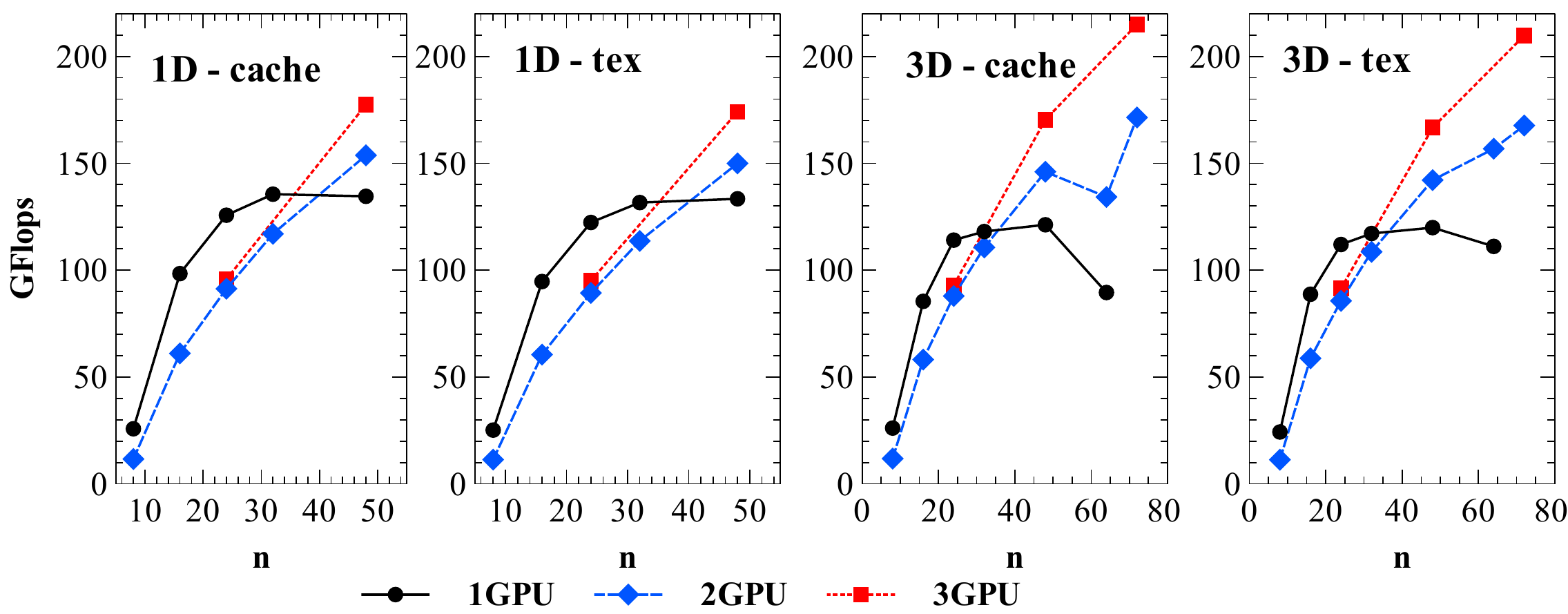}
\par\end{centering}}

    \subfloat[\label{fig:mgpu_su2_580_dp}]{
\begin{centering}
    \includegraphics[height=5cm]{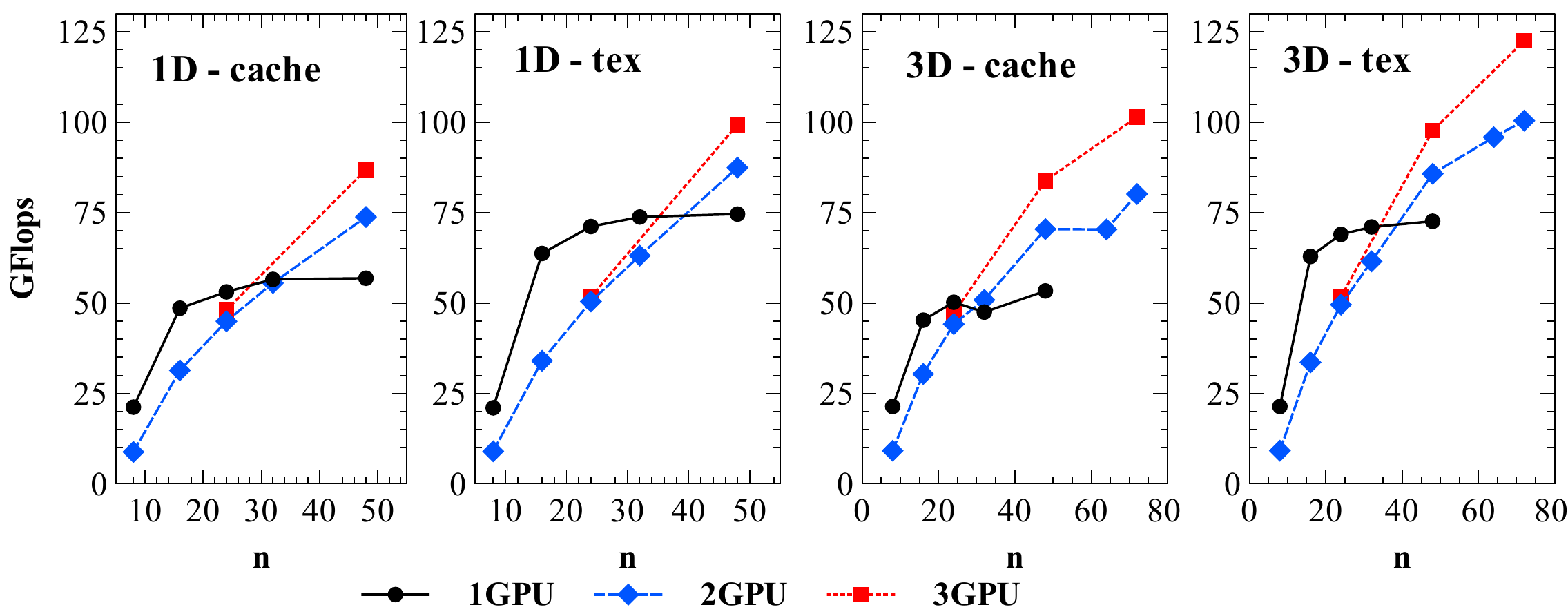}
\par\end{centering}}
\par\end{centering}
    \caption{SU(2) CUDA performance in GFlops using one, two and three GTX 580 GPUs with CUDA and OpenMP. Figs. \protect\subref{fig:mgpu_su2_580_sp} and \protect\subref{fig:mgpu_su2_580_dp} correspond to single and double precision respectively. "n" corresponds to the number of points in each lattice dimension, i.e., $n=N_x=N_y=N_z=N_t$.}
    \label{fig:mgpu_su2_580}
\end{figure}

In Fig. \ref{fig:mgpu_su2_580}, we show the overall performance of the SU(2) code using one, two and three GPUs, NVIDIA GTX 580. 
The performance in multi-GPU mode is dependent on the memory traffic between GPUs and CPU. The memory bandwidth between GPU and CPU is less than the global memory access. Therefore, the performance with more than one GPU is dependent on the lattice size.
In Figs. \ref{fig:scal_su2_sp} and \ref{fig:scal_su2_dp} we plot the scaling performance from one to three GPUs for a fixed lattice size of $48^4$ in single and double precision respectively.

For a $72^4$ lattice volume and using three GPUs, we obtain 210 and 122 GFlops in single and double precision respectively.
Note that we have not yet implemented an overlap between computing and memory transfers and therefore we still can improve the performance with a full overlap in a future implementation.

The implementation with 1D thread blocks has an overall performance higher than the implementation with 3D thread blocks in single precision, but in double precision the performance is practically the same.
Therefore, in the next two subsections, SU(3) and SU(4) CUDA performance, we  test the performance  using 3D thread blocks.

\subsection{SU(3) CUDA performance}

\begin{figure}[!htb]
\begin{centering}
    \includegraphics[width=14cm]{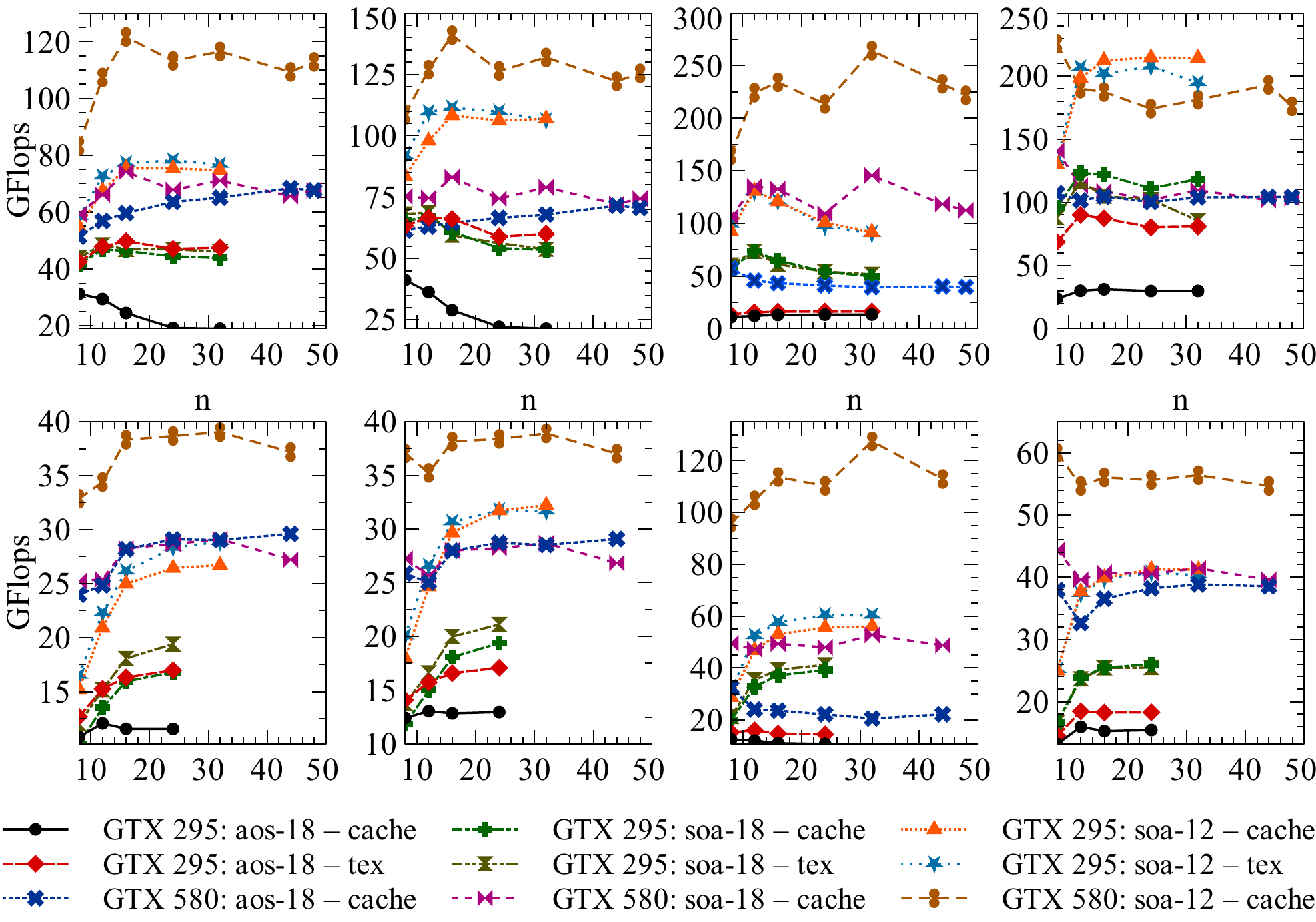}
\par\end{centering}
\caption{SU(3) CUDA performance in GFlops. Left to right: heatbath, over-relaxation, re-unitarization and plaquette kernels. The top line of graphics corresponds to performance in single precision and the bottom to double precision. "n" corresponds to the number of points in each lattice dimension, i.e., $n=N_x=N_y=N_z=N_t$.}
\label{fig:su3_perf}
\end{figure}

We test the performance using three different implementations, a lattice array with 18 real numbers in an array of structures and a structure of arrays and a lattice array with 12 real numbers in a structure of arrays. We also tested the performance in single and double precision for each kernel, Fig. \ref{fig:su3_perf}. In the SU(3) code, the performance is directly affected by the memory transfers and matrix-matrix multiplications. Compared with SU(2), the memory size transfers increase by a factor of 3 and 4.5 for 12 and 18 real number representation in a lattice array and the process to update an SU(3) link consists of three steps of SU(2). 

\begin{figure}[!htb]
\begin{centering}
    \subfloat[\label{fig:mgpu_su3_580_sp}]{
\begin{centering}
    \includegraphics[height=5cm]{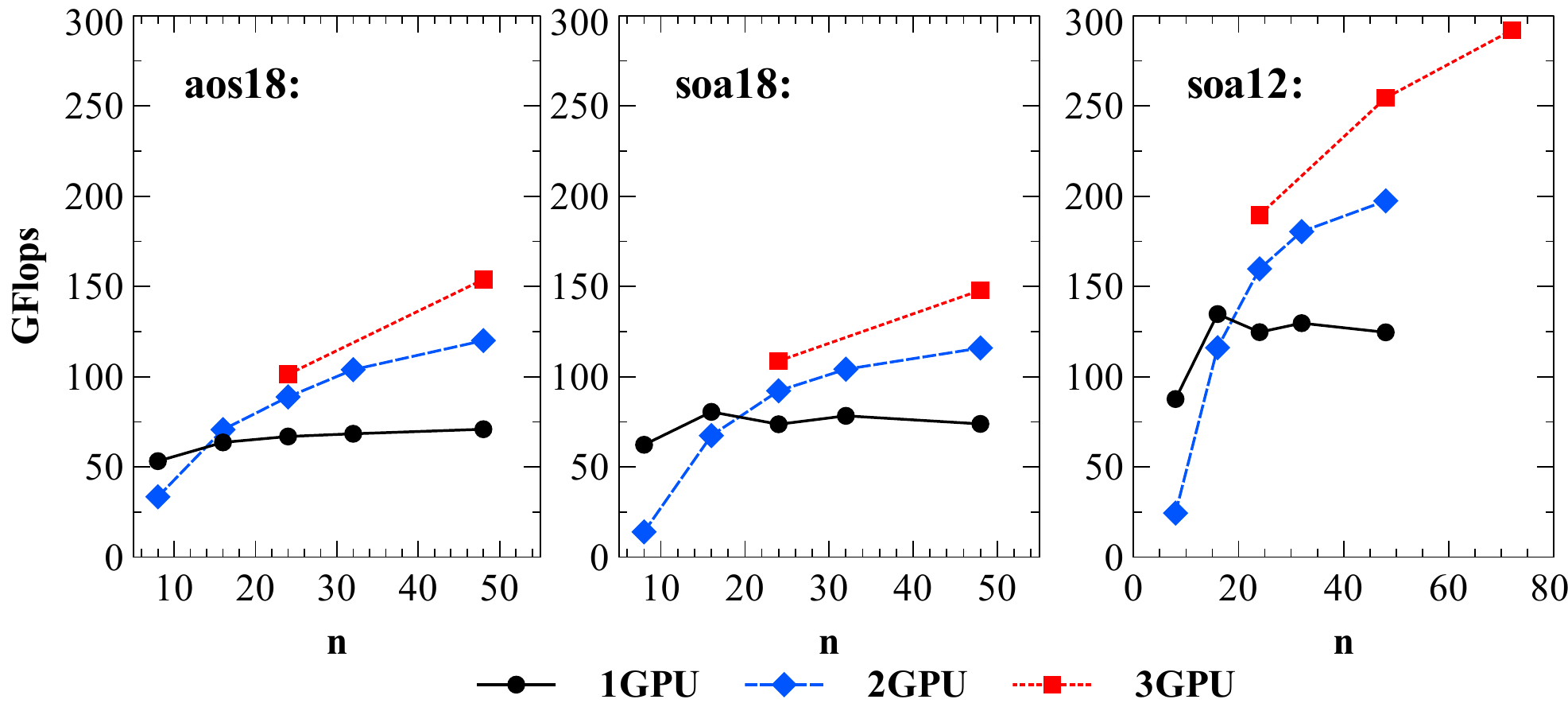}
\par\end{centering}}

    \subfloat[\label{fig:mgpu_su3_580_dp}]{
\begin{centering}
    \includegraphics[height=5cm]{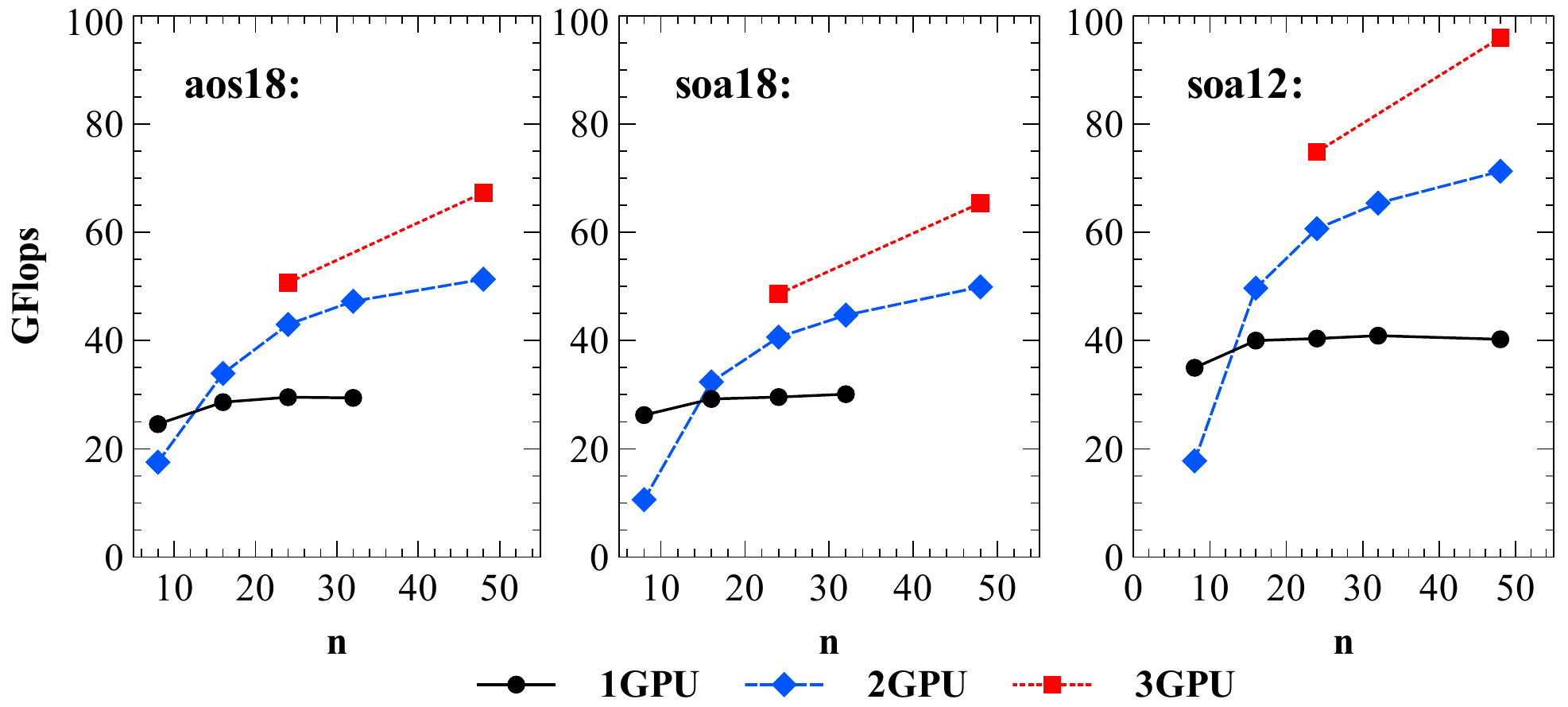}
\par\end{centering}}
\par\end{centering}
    \caption{SU(3) CUDA performance in GFlops using one, two and three GTX 580 GPUs with CUDA and OpenMP. Figs. \protect\subref{fig:mgpu_su3_580_sp} and \protect\subref{fig:mgpu_su3_580_dp} correspond to single and double precision respectively. "n" corresponds to the number of points in each lattice dimension, i.e., $n=N_x=N_y=N_z=N_t$.}
    \label{fig:mgpu_su3_580}
\end{figure}

In Fig. \ref{fig:mgpu_su3_580}, we show the SU(3) performance in GFlops using one, two and three GPUs. As expected, if we reduce the thread memory access from 18 real numbers to 12 real numbers per link, we can increase the performance $1.9\times$ and $1.45\times$ for single and double precision respectively, using three GPUs. 
In single GPU mode the best way to store the full lattice is a SOA. However, in multi-GPU mode the AOS implementation is better.
 In the AOS implementation the number of copies is less than the number of copies in the SOA implementation, while the amount of memory size transferred is the same.
In Figs. \ref{fig:scal_su3_sp} and \ref{fig:scal_su3_dp} we plot the scaling performance from one to three GPUs for a fixed lattice size of $48^4$ in single and double precision respectively.
Reducing the size of memory transfers, in this case from 18 to 12 real numbers, the performance increases. Another important aspect of using 12 real numbers is that we can have bigger lattice sizes. 
 Therefore, using one or more GPUs is intrinsically dependent on how the data is organized in the memory.

For a $72^4$ lattice volume and using three GPUs, we obtain 292 and 96 GFlops in single and double precision respectively.

\subsection{SU(4) CUDA performance}

In this subsection, we test the SU(4) CUDA code performance using an array of structures and a structure of arrays in single and double precision. As in the previous subsection, we only implemented and tested 3D thread blocks. 

\begin{figure}[!htb]
\begin{centering}
    \includegraphics[width=14cm]{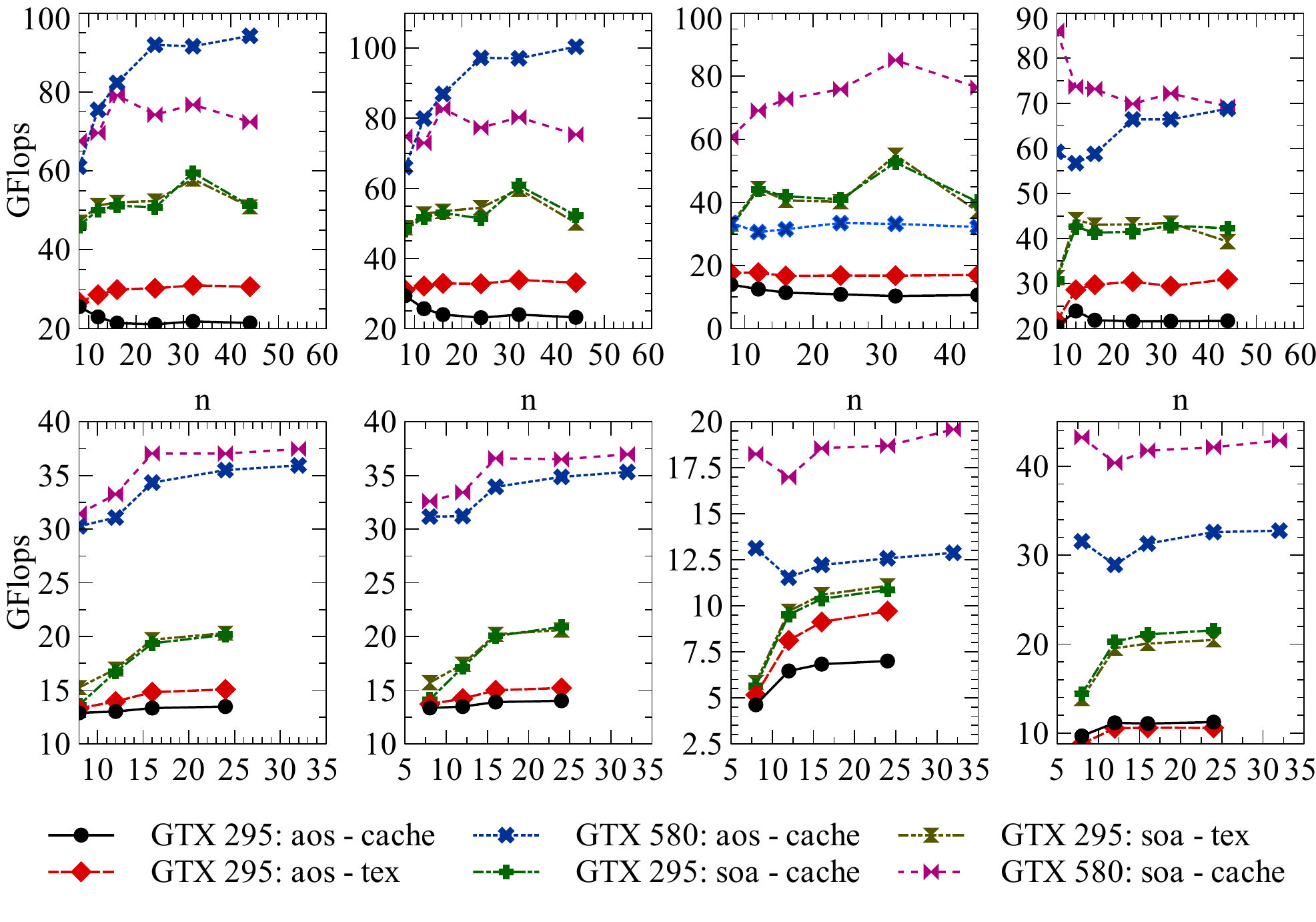}
\par\end{centering}
\caption{SU(4) CUDA performance in GFlops. Left to right: heatbath, over-relaxation, re-unitarization and plaquette kernels. The top line of graphics corresponds to performance in single precision and the bottom to double precision. "n" corresponds to the number of points in each lattice dimension, i.e., $n=N_x=N_y=N_z=N_t$.}
\label{fig:su4_perf}
\end{figure}

In Fig. \ref{fig:su4_perf}, we show performance in GFlops as a function of the lattice volume, for each kernel in single GPU mode. Compared with SU(2), the process to update an SU(4) link consists of six steps of SU(2), three steps more than SU(3).
In SU(4), the memory size increased eight times compared with the SU(2) code.

\begin{figure}[!htb]
\begin{centering}
    \subfloat[\label{fig:mgpu_su4_580_sp}]{
\begin{centering}
    \includegraphics[height=5cm]{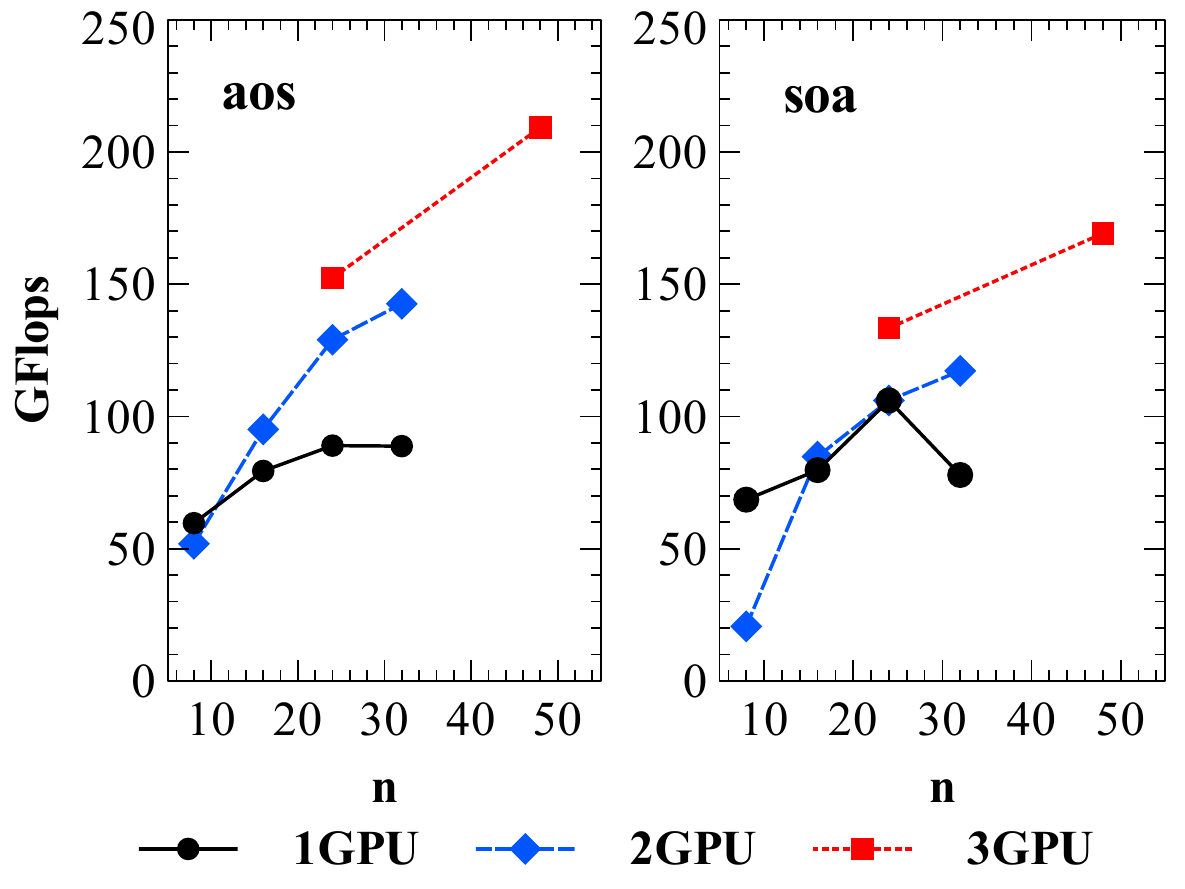}
\par\end{centering}}
    \subfloat[\label{fig:mgpu_su4_580_dp}]{
\begin{centering}
    \includegraphics[height=5cm]{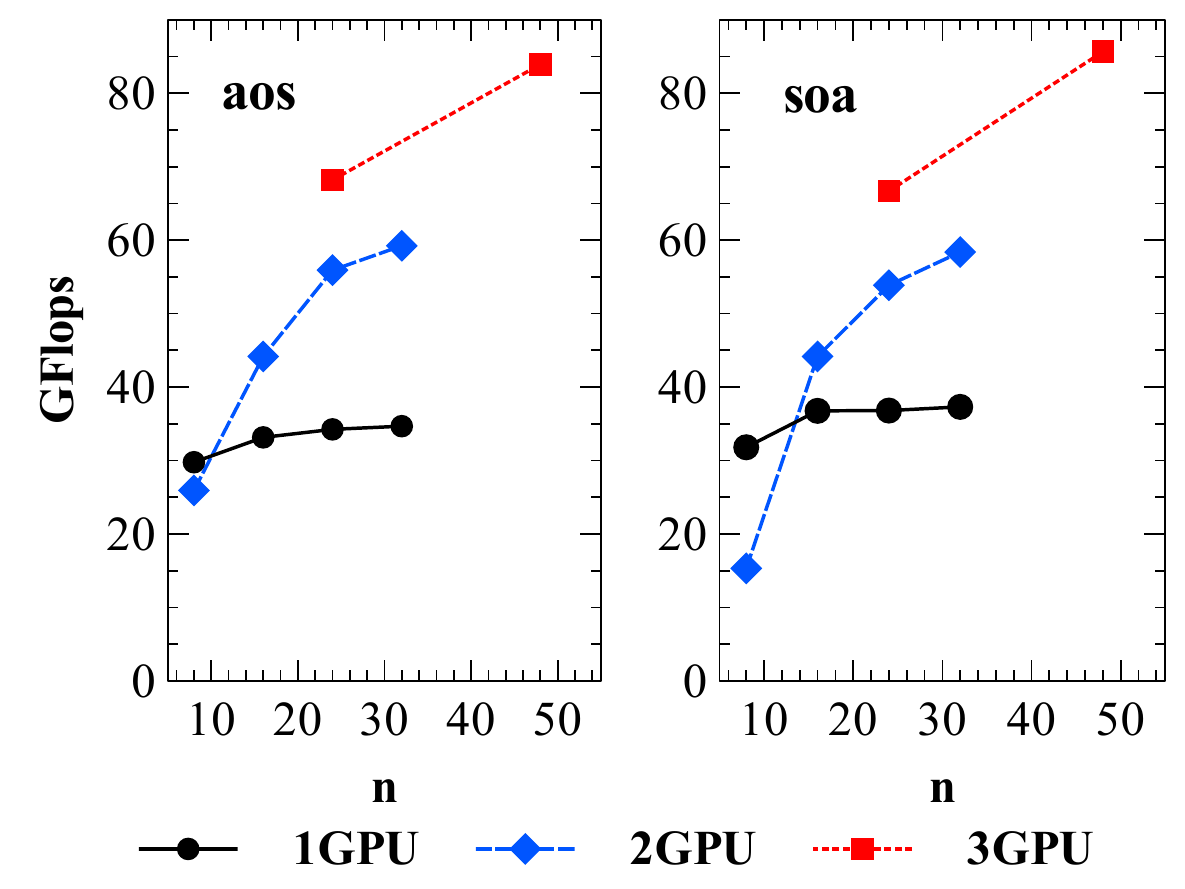}
\par\end{centering}}
\par\end{centering}
    \caption{SU(4) CUDA performance in GFlops using one, two and three GTX 580 GPUs with CUDA and OpenMP. Figs. \protect\subref{fig:mgpu_su4_580_sp} and \protect\subref{fig:mgpu_su4_580_dp} correspond to single and double precision respectively. "n" corresponds to the number of points in each lattice dimension, i.e., $n=N_x=N_y=N_z=N_t$.}
    \label{fig:mgpu_su4_580}
\end{figure}

\begin{figure}[!htb]
\begin{centering}
    \subfloat[SU(2) single precision.\label{fig:scal_su2_sp}]{
\begin{centering}
    \includegraphics[height=5.4cm]{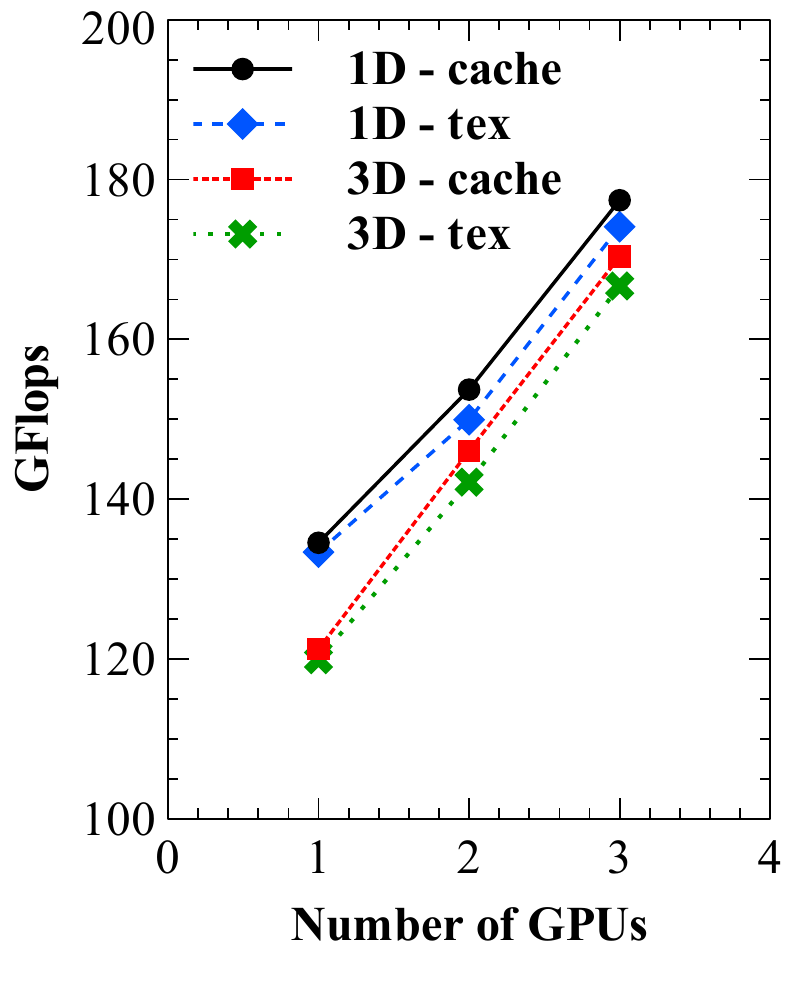}
\par\end{centering}}
    \subfloat[SU(3) single precision.\label{fig:scal_su3_sp}]{
\begin{centering}
    \includegraphics[height=5.4cm]{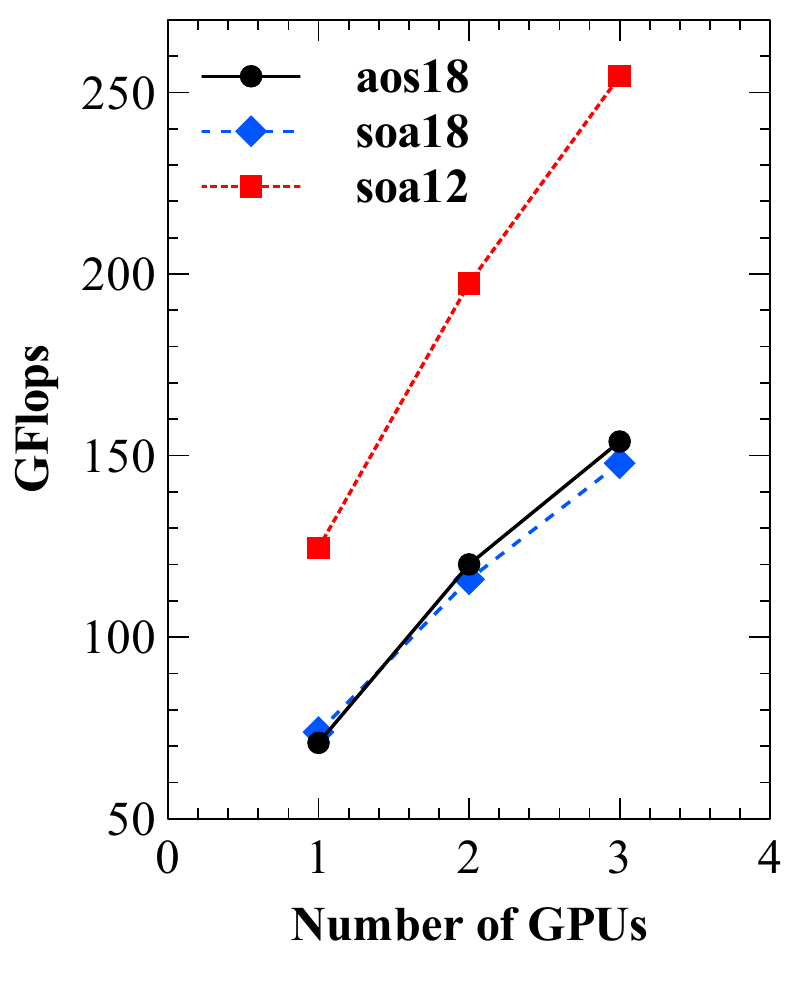}
\par\end{centering}}
    \subfloat[SU(4) single precision.\label{fig:scal_su4_sp}]{
\begin{centering}
    \includegraphics[height=5.4cm]{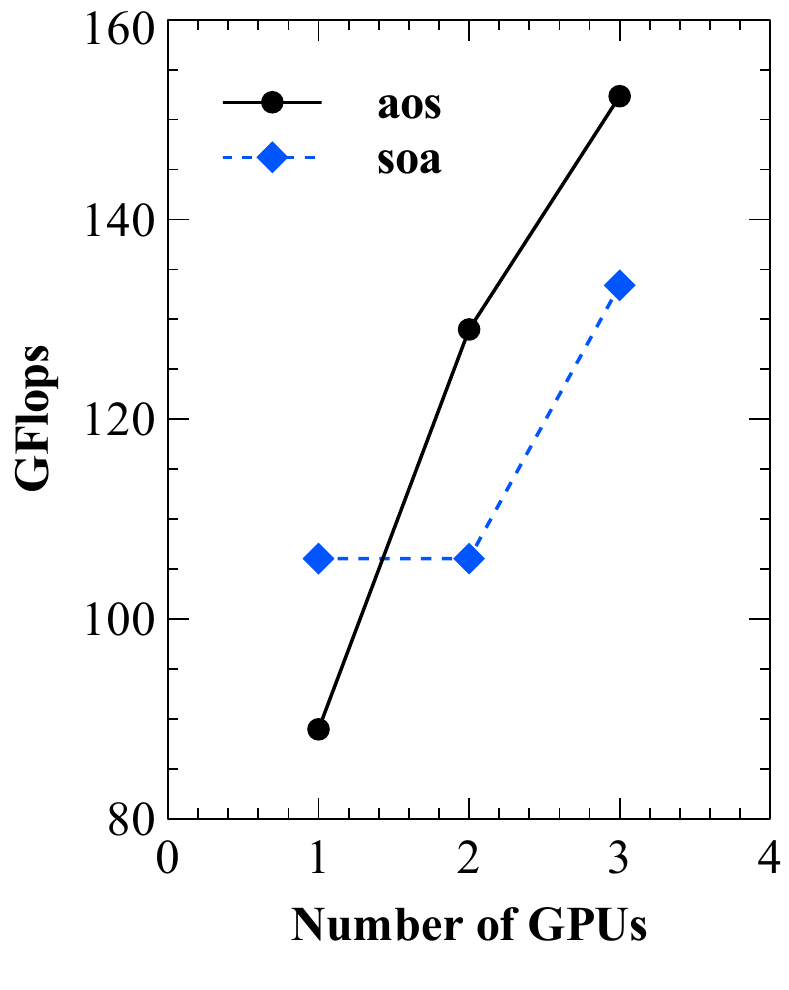}
\par\end{centering}}

    \subfloat[SU(2) double precision.\label{fig:scal_su2_dp}]{
\begin{centering}
    \includegraphics[height=5.4cm]{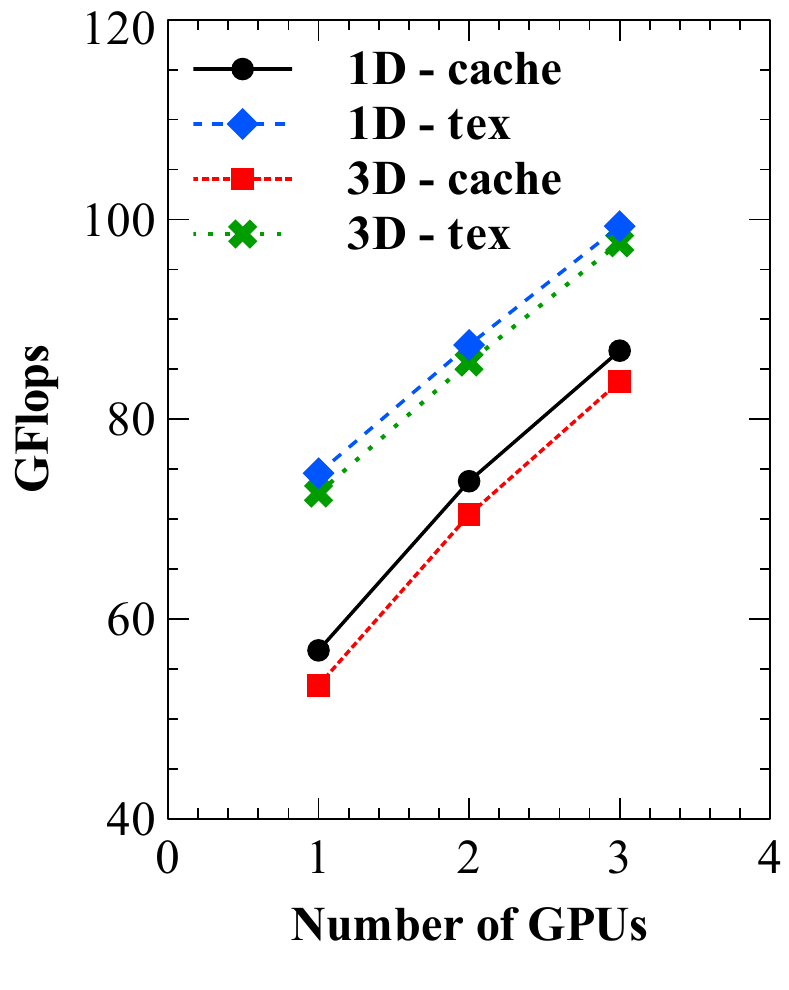}
\par\end{centering}}
    \subfloat[SU(3) double precision.\label{fig:scal_su3_dp}]{
\begin{centering}
    \includegraphics[height=5.4cm]{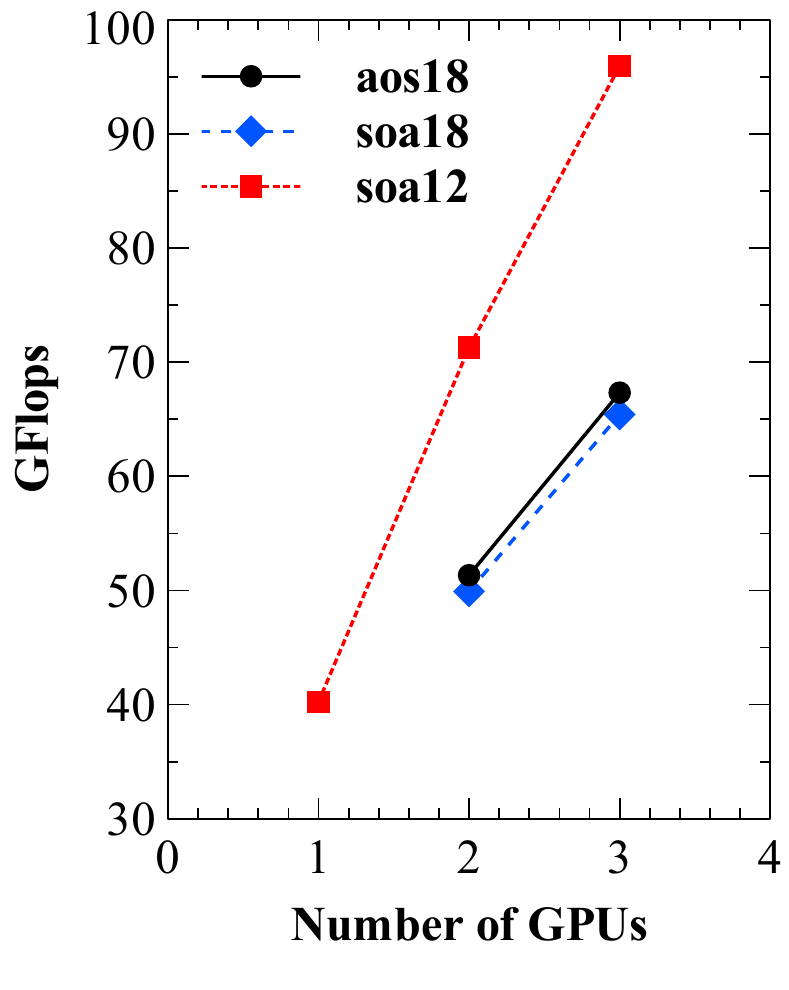}
\par\end{centering}}
    \subfloat[SU(4) double precision.\label{fig:scal_su4_dp}]{
\begin{centering}
    \includegraphics[height=5.4cm]{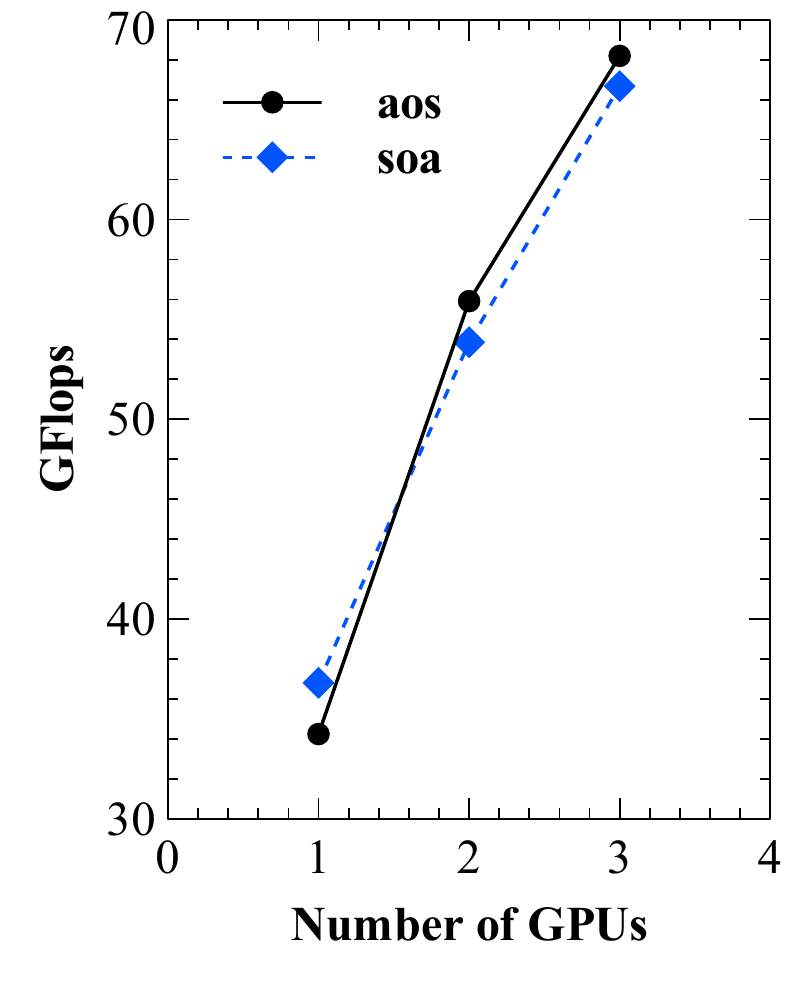}
\par\end{centering}}
\par\end{centering}
    \caption{SU(2), SU(3) and SU(4) CUDA performance in GFlops as a function of the number of GPUs in single and double precision. The lattice volume is kept constant at $48^4$ for SU(2) and SU(3) and $24^4$ for SU(4).}
    \label{fig:scal_sun_580}
\end{figure}

In Fig. \ref{fig:mgpu_su4_580} we show the performance in GFlops for one, two and three GPUs, in single and double precision. The AOS implementation gives better results in single precision. However, in double precision there is almost no difference between the AOS and SOA implementations.
In Figs. \ref{fig:scal_su4_sp} and \ref{fig:scal_su4_dp} we plot the scaling performance from one to three GPUs for a fixed lattice size of $24^4$ in single and double precision respectively. Therefore, using one or more GPUs is intrinsically dependent on how the data is organized in the memory.

For a $48^4$ lattice volume and using three GPUs, we obtain 169 and 86 GFlops in single and double precision respectively.

\subsection{Generic SU($N_c$) CUDA performance}

\begin{figure}[!h]
\begin{centering}
    \includegraphics[width=11cm]{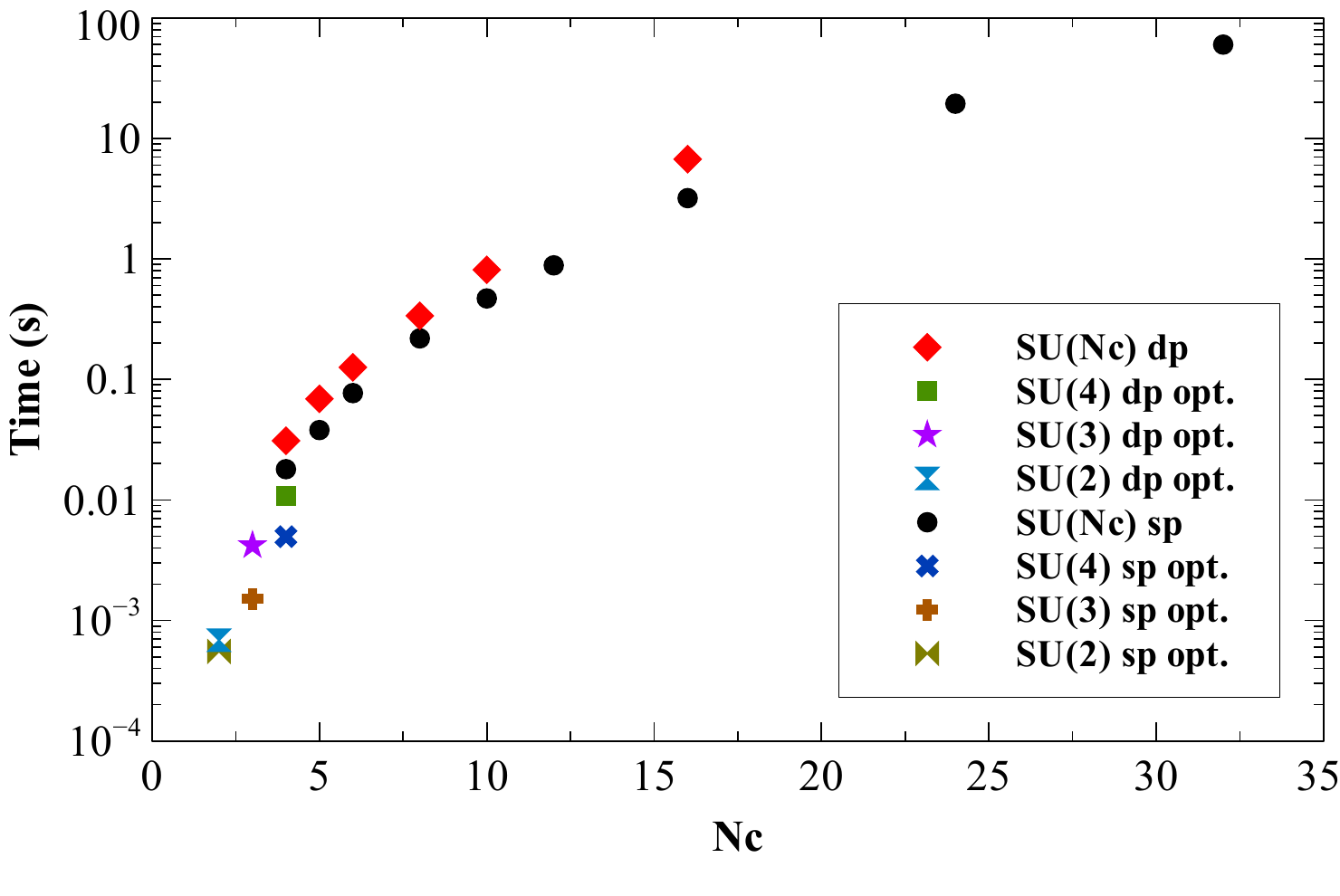}
\par\end{centering}
\caption{SU($N_c$) CUDA time per iteration in single (sp) and double (dp) precision for a $8^4$ lattice volume and $\beta=6.2$ running in a GTX 580 GPU. One iteration consists of one pseudo-heatbath step, one over-relaxation step followed by link re-unitarization. Comparison between the generic SU($N_c$) code for $N_c\,\geq 4$ (black circle and red diamond) with the optimized SU(2), SU(3) and SU(4) codes.}
\label{fig:sun_perf}
\end{figure}

In Fig. \ref{fig:sun_perf} we compare the time to perform one iteration in the SU(2), SU(3), SU(4) optimized codes and a generic code valid for $N_c \ge 4$. As expected, the SU(4) generic code is less efficient than the optimized SU(4) code. 

Thus the application of our code to very large $N_c$ is significantly slower than the small $N_c$ codes due to the optimization and to the scaling with $N_c$ of the generic code.  
For the same lattice size, generating configurations in SU(32) is six orders of magnitude slower that in SU(2).

In what concerns our generic code valid only for $N_c \geq 4$, as $N_c$ increases, the memory accesses also increase, and the number of subgroups of SU(2) increase with $N_c(N_c-1)/2$ as well as the number of floating point operations, since each link is made of $N_c\times N_c\times 2$ floating numbers.

As seen previously in the SU(2), SU(3) and SU(4) performance results, the highest performance depends intrinsically on the group number, $N_c$, on how the data is organized in GPU memory and on the number of GPUs. Thus, to get the highest performance for generic $N_c$'s, especially for large ones, this is a highly demanding task, due to the limited GPU resources compared to the CPU, namely memory and registers.
In the single precision case we can go up to $N_c=32$ and in double precision we can go up to $N_c=16$ for a lattice volume $8^4$ using a GTX 580 GPU.

\begin{figure}[!h]
\begin{centering}
    \includegraphics[width=11cm]{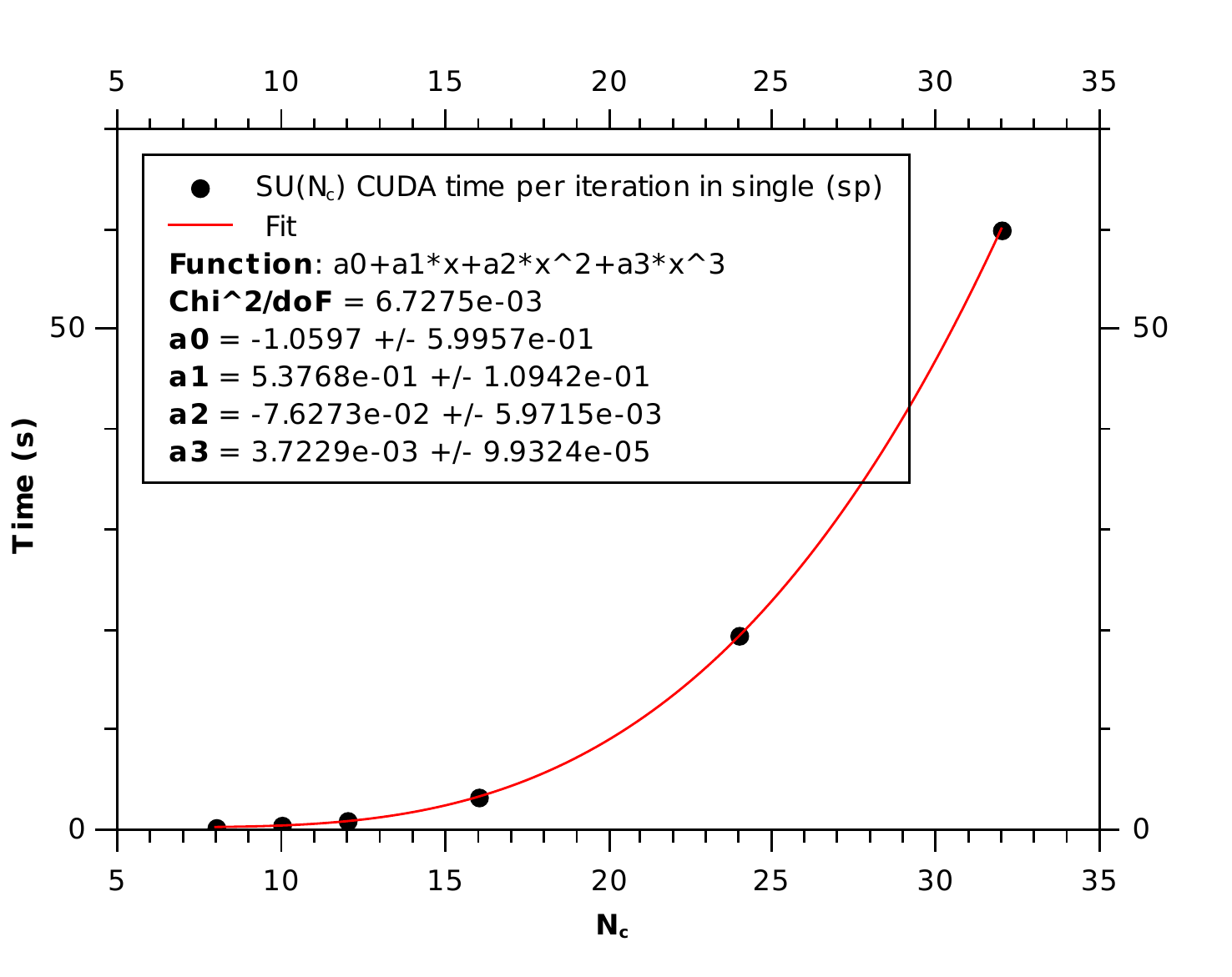}
\par\end{centering}
\caption{Fit of the SU($N_c$) CUDA time (s) per iteration in single precision, shown in Fig. \ref{fig:sun_perf}. The polynomial fit up to ${N_c}^3$ agrees with scaling of the pseudo-heatbath/overrelaxation steps.}
\label{fig:sun_fit}
\end{figure}

In our generic code, the time to perform one iteration goes up as the third power of $N_c$, Fig. \ref{fig:sun_fit}. This agrees with the fact that to perform a pseudo-heatbath/overrelaxation step there are $N_c(N_c-1)/2$ subgroups of SU(2) to be generated and then multiplied by the actual link.
Each multiplication is of the order of $N_c$ and therefore this gives the $N_c^3$ factor.
Note that the calculation of the staple, needed in the pseudo-heatbath/overrelaxation steps, also goes with $N_c^3$.

\section{Conclusion}

We developed codes in CUDA to generate pure gauge SU($N_c$) configurations for lattice QCD simulations. 

The technique used to store the SU($N_c$) elements in the global memory is important to achieve the best performance.  We produce specific codes for SU(2), SU(3) and SU(4), to optimize the group parameterizations. We also implement a generic code valid for any SU($N_c$), taking into account the best way to store the elements in global memory. 

Due to the limited amount of GPU resources per thread, because the matrix size is $N_c\times N_c$, the implementation of the generic SU($N_c$) cannot operate for an arbitrarily  large $N_c$.  
As $N_c$ increases, the memory accesses also increase, and the number of subgroups of SU(2) increases with $N_c(N_c-1)/2$ as well as the number of floating point operations, since each link is made of $N_c\times N_c\times 2$ floating numbers. Thus the codes with very large $N_c$ are significantly slower than those with moderate $N_c$ ones. 

Nevertheless, as shown in Fig. \ref{fig:sun_perf}, up to SU(6) our  generic SU($N_c$) configuration generation code is only one order of magnitude slower than the optimized SU(3) code, and it is possible up to some extent to study the large $N_c$ limit.

\section*{Acknowledgments}
This work was partly funded by the FCT contracts, POCI/FP/81933/2007,
CERN/FP/83582/2008, PTDC/FIS/100968/2008, CERN/FP/109327/2009
and CERN/FP/116383/2010.

Nuno Cardoso is also supported by FCT under the contract SFRH/BD/44416/2008.

We thank Marco Cardoso for useful discussions.
\bibliographystyle{elsarticle-num}
\bibliography{bib}

\end{document}